\documentclass[preprint,aps,prd,amsmath,amssymb]{revtex4}
\usepackage{graphicx}
\usepackage{color}
\usepackage{graphicx}
\usepackage{dcolumn}
\usepackage{bm}
\usepackage{amsmath}
\usepackage{amsfonts}
\usepackage{bbm}
\usepackage{subfigure}
\usepackage{setspace}
\newcommand{\beq}{\begin{eqnarray}}
\newcommand{\eeq}{\end{eqnarray}}

\newcommand{\bmp}{\noindent\begin{minipage}{16cm}}
\newcommand{\emp}{\end{minipage}\vskip 7mm} 

\def\drawbox#1#2{\hrule height#2pt
        \hbox{\vrule width#2pt height#1pt \kern#1pt
              \vrule width#2pt}
              \hrule height#2pt}

\def\Asym#1#2{\vcenter{\vbox{\drawbox{#1}{#2}
              \kern-#2pt 
              \drawbox{#1}{#2}}}}

\begin{document}
\title{\Large  Unitarity in Technicolor}
\author{Roshan {\sc Foadi}}
\email{roshan@ifk.sdu.dk}
\author{Matti  {\sc J\"arvinen}}
\email{mjarvine@ifk.sdu.dk}
\author{Francesco {\sc Sannino}}
\email{sannino@ifk.sdu.dk}
\affiliation{Centre for High Energy Physics, University of Southern Denmark \\ Campusvej 55, DK-5230 Odense M, Denmark.}


\begin{abstract}
We investigate the longitudinal $WW$ scattering in models of dynamical electroweak symmetry breaking featuring a spin one axial and vector state and a composite Higgs. We also investigate the effects of a composite spin two state which has the same properties of a massive graviton. Any model of dynamical electroweak symmetry breaking will feature, depending on the dynamics, some or all these basic resonances as part of the low energy spectrum. We suggest how to take limits in the effective Lagrangian parameter space to reproduce the dynamics of different types of underlying gauge theories, from the traditional Technicolor models to the newest ones featuring nearly conformal dynamics. We study the direct effects of a light composite Higgs and the indirect ones stemming from the presence of a light axial resonance on the longitudinal $WW$ scattering. 
\end{abstract}

\maketitle

\section{Introduction}

The Standard Model (SM) constitutes without any doubt one of the  most successful models of Nature. Despite such an astounding success the SM sector describing the breaking of the electroweak symmetry has not been experimentally confirmed. In fact, there is a fair chance that it might be described by a novel strongly coupled dynamics~\cite{Sannino:2008ha} inspired to the old Technicolor models~\cite{Weinberg:1979bn,Susskind:1978ms}. 

Precision data, as well as flavor changing neutral currents constraints (FCNC) require the new strong dynamics to be different from the QCD one. Nearly conformal technicolor models can simultaneously reduce the tension with precision data~\cite{Appelquist:1998xf,Kurachi:2006mu} and suppress  dangerous FCNC~\cite{Eichten:1979ah,Holdom:1981rm,Yamawaki:1985zg,Appelquist:1986an}. 

In order to be prepared for such a discovery at the Large Hadron Collider we have introduced a few explicit models passing the electroweak precision tests as summarized in~\cite{Sannino:2008ha}. Two examples are Minimal Walking Technicolor (MWT)~\cite{Sannino:2004qp,Dietrich:2005jn,Dietrich:2006cm,Foadi:2007ue} and Ultra Minimal Technicolor (UMT)~\cite{Ryttov:2008xe}. The models constitute interesting benchmarks for collider phenomenology~\cite{Foadi:2007ue,Belyaev:2008yj}. Moreover MWT, with additional adjoint SM fermions, leads to the unification of the SM couplings~\cite{Gudnason:2006mk} and to even new candidates of cold dark matter type~\cite{Gudnason:2006yj,Kouvaris:2007iq,Kainulainen:2006wq,Khlopov:2008ty,Kouvaris:2008hc}. UMT phenomenology is very rich although its collider signals remain to be explored. It features a novel intriguing candidate for cold Dark Matter, the Technicolor Interacting Massive particle (TIMP). The TIMP  is identified with a pseudo Goldstone technibaryon. Another relevant fact is that these models have the potential to explain baryogenesis since they can lead to a first order electroweak phase transition as a function of the temperature~\cite{Cline:2008hr}. 

To construct these models we used recent explorations of the phase diagram of strongly coupled gauge theories as a function of  the number of colors, flavors, and matter representation. We combined novel~\cite{Ryttov:2007cx,Ryttov:2007sr} and older analytic methods~\cite{Sannino:2004qp,Dietrich:2006cm} together with recent first principles Lattice simulations~\cite{Catterall:2007yx,Catterall:2008qk,Shamir:2008pb,DelDebbio:2008wb,DelDebbio:2008zf,Hietanen:2008vc,Appelquist:2007hu,Deuzeman:2008sc,Fodor:2008hn}.

An essential point, which was first made in~\cite{Hong:2004td} and then in~\cite{Dietrich:2005jn}, is that these models may feature a light composite Higgs (LCH). In Appendix F of~\cite{Sannino:2008ha} one of the authors has shown, using the Corrigan and Ramond large N limit of QCD~\cite{Corrigan:1979xf}, how a LCH naturally emerges in a strongly coupled theory with higher dimensional representations. Near conformal dynamics can further help keeping this state light relative to the electroweak scale even at a small number of colors~\cite{Sannino:2008ha,Dietrich:2005jn} . 

The spin one sector is also very interesting. Thanks to the nearly conformal dynamics the second Weinberg Sum Rule (WSR) is modified~\cite{Appelquist:1998xf}. This allows for the first spin one axial resonance to be lighter than the vector one. It is then interesting to investigate the effect of a LCH and a light axial resonance (LAR) on the longitudinal $WW$ scattering amplitude. A systematic study of the collider phenomenology of a LCH and a LAR at the LHC has begun in Ref~\cite{Belyaev:2008yj}, where it is shown that the associate Higgs production together with a SM vector boson is one of the interesting signals.  

We also investigate the effect of a massive spin-two resonance on the longitudinal $WW$ scattering. This is also relevant since an isosinglet massive spin two particle  may very well be mis-identified as a massive graviton  stemming from a less natural extra dimensional extension of the SM.  

The analysis presented here generalizes the results of~\cite{Foadi:2008ci} by adding the LCH and the spin-two state. The present analysis is valid when the resonance exchanges dominate the dynamics.  It is, in practice, the principle of vector meson dominance (VMD). Differently from QCD~\cite{Sannino:1995ik,Harada:1995dc} we have a {\it narrow} light composite scalar (the Higgs). Loop corrections can be investigated, however VMD is expected to be an efficient way to take into account these corrections. 


\section{Unitarity of Pion Pion Scattering in Technicolor}

Consider a strongly interacting gauge theory with an SU(2)$_{\rm L}\times$SU(2)$_{\rm R}$ chiral symmetry. Suppose this new strong interaction spontaneously breaks the chiral symmetry to SU(2)$_{\rm V}$. If we identify the electroweak gauge group with the SU(2)$_{\rm L}\times$U(1)$_{\rm R}$ subgroup of SU(2)$_{\rm L}\times$SU(2)$_{\rm R}$ this becomes a model of Technicolor. At low energy, below the confining scale, this theory is described by an effective Lagrangian in which the bound states can be classified according to the chiral symmetry group. 

In the effective theory the scattering amplitudes for the longitudinal SM gauge bosons approach at large energies the scattering amplitudes for the corresponding eaten pions. We mainly analyze the contribution to the $\pi\pi$ scattering amplitude from a spin-zero isosinglet and a spin-one isotriplet, and consider the case in which a spin-two isosinglet contributes as well.

\subsection{Spin-Zero + Spin-One}

If a spin-zero isosinglet $H$ and a spin-one isotriplet $V^a_\mu$ are in the low energy spectrum, the ${\cal O}(p^2)$ Lagrangian terms contributing to the tree-level pion scattering amplitudes are
\begin{eqnarray}
{\cal L}_{V\pi\pi} & = & g_{V\pi\pi}\ \varepsilon^{abc}\ V^a_\mu\ \pi^b\ \partial^\mu\pi^c \ , \label{eq:LVpp_main} \\
{\cal L}_{H\pi\pi} & = & h_1\ M_H\ H\ \pi^a\ \pi^a + \frac{h_2}{F_\pi}\ H\ \partial^\mu\pi^a\ \partial_\mu\pi^a
+ \frac{h_3}{F_\pi}\ \partial^\mu H \partial_\mu\pi^a\ \pi^a \ , \label{eq:LHpp_main} \\
{\cal L}_{\pi\pi\pi\pi} & = & g_1\ \pi^a\ \pi^a\ \pi^b\ \pi^b + \frac{g_2}{F_\pi^2}\ \pi^a\ \pi^a\ \partial^\mu\pi^b\ \partial_\mu\pi^b
+ \frac{g_3}{F_\pi^2}\ \pi^a\ \partial^\mu\pi^a\ \pi^b\ \partial_\mu\pi^b \ , \label{eq:Lpppp_main}
\end{eqnarray}
where $F_\pi$ is the pion decay constant. Since this is a model of Technicolor, $F_\pi\simeq 246$ GeV. $V^a_\mu$ is a parity-odd spin-one resonance, analog to the QCD $\rho$ meson, while $H$ is a composite Higgs. Notice that our normalization for $g_{V\pi\pi}$ differs by a factor of $\sqrt{2}$ from that of Ref.~\cite{Harada:1995dc}.

The isospin invariant amplitude for the pion-pion elastic scattering is
\begin{equation}
A(s,t,u)=8 g_1 +2(g_3-2g_2)\frac{s}{F_\pi^2}
-\frac{\left[2 M_H h_1+(h_3-h_2)s/F_\pi\right]^2}{s-M_H^2}
-g_{V\pi\pi}^2\left[\frac{s-u}{t-M_V^2}+\frac{s-t}{u-M_V^2}\right] \ .
\label{eq:inv_1}
\end{equation}
Notice that in the way they are written the Lagrangian terms of Eqs.~(\ref{eq:LVpp_main})$\div$(\ref{eq:Lpppp_main}) are only invariant under the unbroken SU(2)$_{\rm V}$ symmetry, with the pions and the vector transforming as triplets, and the Higgs as a singlet of SU(2)$_{\rm V}$. This implies that the corresponding couplings are unrelated. However, as explicitly shown in Appendix~\ref{appA}, in our approach $H$, $\pi^a$, and $V^a_\mu$ do indeed transform under the full SU(2)$_{\rm L}\times$SU(2)$_{\rm R}$ chiral symmetry, which spontaneously breaks to the isospin symmetry SU(2)$_{\rm V}$. This implies the relations
\begin{eqnarray}
& & g_1 = -\frac{h_1^2}{2} \\
& & \frac{8g_1}{M_H^2}+\frac{4 h_1(h_2-h_3)}{M_H F_\pi}-\frac{2(g_3-2g_2)}{F_\pi^2}=-\frac{1}{F_\pi^2}+\frac{3 g_{V\pi\pi}^2}{M_V^2} \ ,
\end{eqnarray}
which are easy to prove by using the formulas in Appendix~\ref{appA}. Inserting this in Eq.~(\ref{eq:inv_1}), and defining
\begin{equation}
h\equiv 2 h_1-\frac{M_H}{F_\pi}(h_2-h_3) \ ,
\label{eq:h}
\end{equation}
leads to
\begin{equation}
A(s,t,u)=\left(\frac{1}{F_\pi^2}-\frac{3g_{V\pi\pi}^2}{M_V^2}\right)s
-\frac{h^2}{M_H^2}\frac{s^2}{s-M_H^2}
-g_{V\pi\pi}^2\left[\frac{s-u}{t-M_V^2}+\frac{s-t}{u-M_V^2}\right] \ ,
\label{eq:inv_2}
\end{equation}
in agreement with the result of Ref.~\cite{Harada:1995dc} for the $\pi\pi$ scattering in QCD. The latter was obtained in a nonlinearly realized effective theory, in which the bound states are classified according to the stability group SU(2)$_{\rm V}$, rather than the full SU(2)$_{\rm L}\times$SU(2)$_{\rm R}$ chiral symmetry group. The two approaches are indeed proven to be equivalent at tree-level.

Notice that the amplitude of Eq.~(\ref{eq:inv_2}) has an $s$-channel pole in the Higgs exchange. In the vicinity of this pole the propagator should be modified to include the Higgs width. In order to catch the essential features of the unitarization process we will take the Higgs to be a relatively narrow state, and consider values of $\sqrt{s}$ far away from $M_H$, where the finite width effects can be neglected. If the Higgs or any other state is not sufficiently narrow to be treated at the tree level, it would be relevant to investigate the effects due to unitarity corrections using specific unitarization schemes as done for example in Ref.~\cite{Black:2000qq}.

\begin{figure}
\includegraphics[width=0.62\textwidth,height=0.47\textwidth]{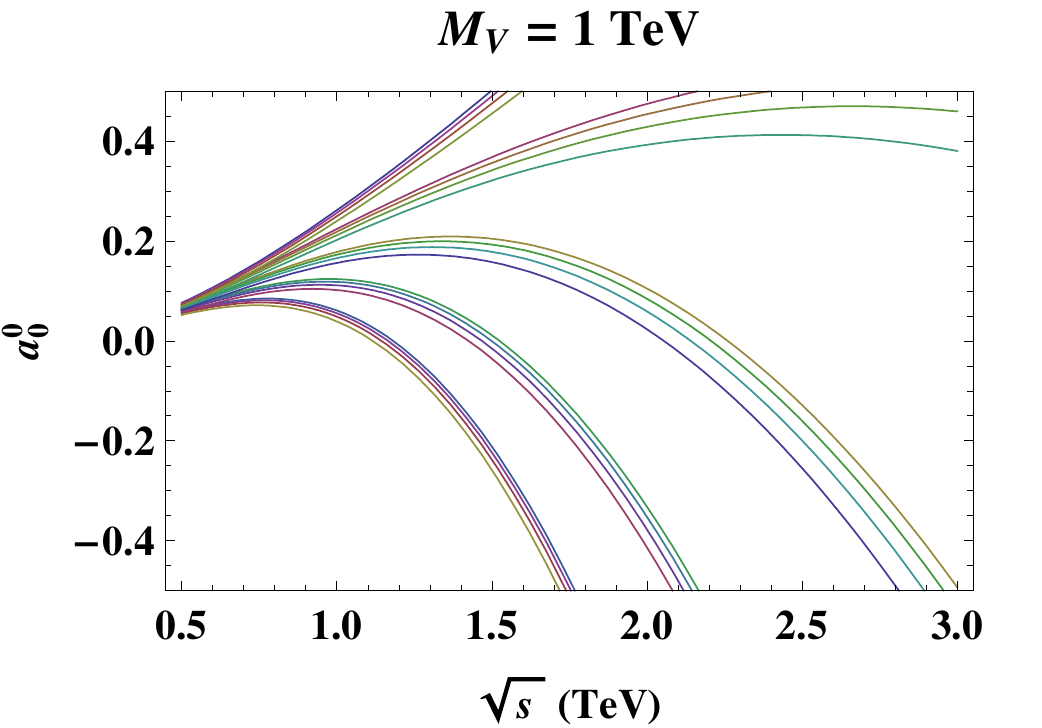}
\caption{$I=0$ $J=0$ partial wave amplitude for the $\pi\pi$ scattering. Here a Higgs with mass $M_H=200$ GeV, and a spin-one vector meson with mass $M_V=1$ TeV contribute to the full amplitude. The different groups of curves correspond, from top to bottom, to $g_{V\pi\pi}=2,2.5,3,3.5,4$. The different curves within each group correspond, from top to bottom, to $h=0,0.1,0.15,0.2$. Nonzero values of $g_{V\pi\pi}$ and $h$ give negative contributions to the linear term in $s$ in the amplitude, and may lead to a delay of unitarity violation.}
\label{fig:a00}
\end{figure}
In order to study unitarity of the $\pi\pi$ scattering the most general amplitude should be expanded in its isospin $I$ and spin $J$ components, $a^I_J$. However the $I=0$ $J=0$ component,
\begin{equation}
a_0^0(s) = \frac{1}{64\pi} \int_{-1}^1 d\cos\theta \left[3A(s,t,u)+A(t,s,u)+A(u,t,s)\right] \ ,
\end{equation}
has the worst high energy behavior, and is therefore sufficient for our analysis. Since we are interested in testing unitarity at few TeV's in presence of a light Higgs, we set
$M_H=200$ GeV as a reference value, and study the regions in the $(M_V,g_{V\pi\pi})$ plane in which $a_0^0$ is unitary up to 3 TeV, for different values of $h$. If the Higgs mass is larger than 200 GeV but still smaller than or of the same size of $M_V$, we expect our results to be qualitatively similar, even though finite width effects might be important due to the pole in the $s$-channel. If the Higgs mass is much larger than $M_V$ the theory is Higgsless at low energies. This case was studied in Ref.~\cite{Foadi:2008ci}, and applies also to the light Higgs scenario if $H$ is decoupled from the pions, {\em i.e.} $h=0$.

\begin{figure}[!t]
\vspace{1.5cm}
\includegraphics[width=0.32\textwidth,height=0.3\textwidth]{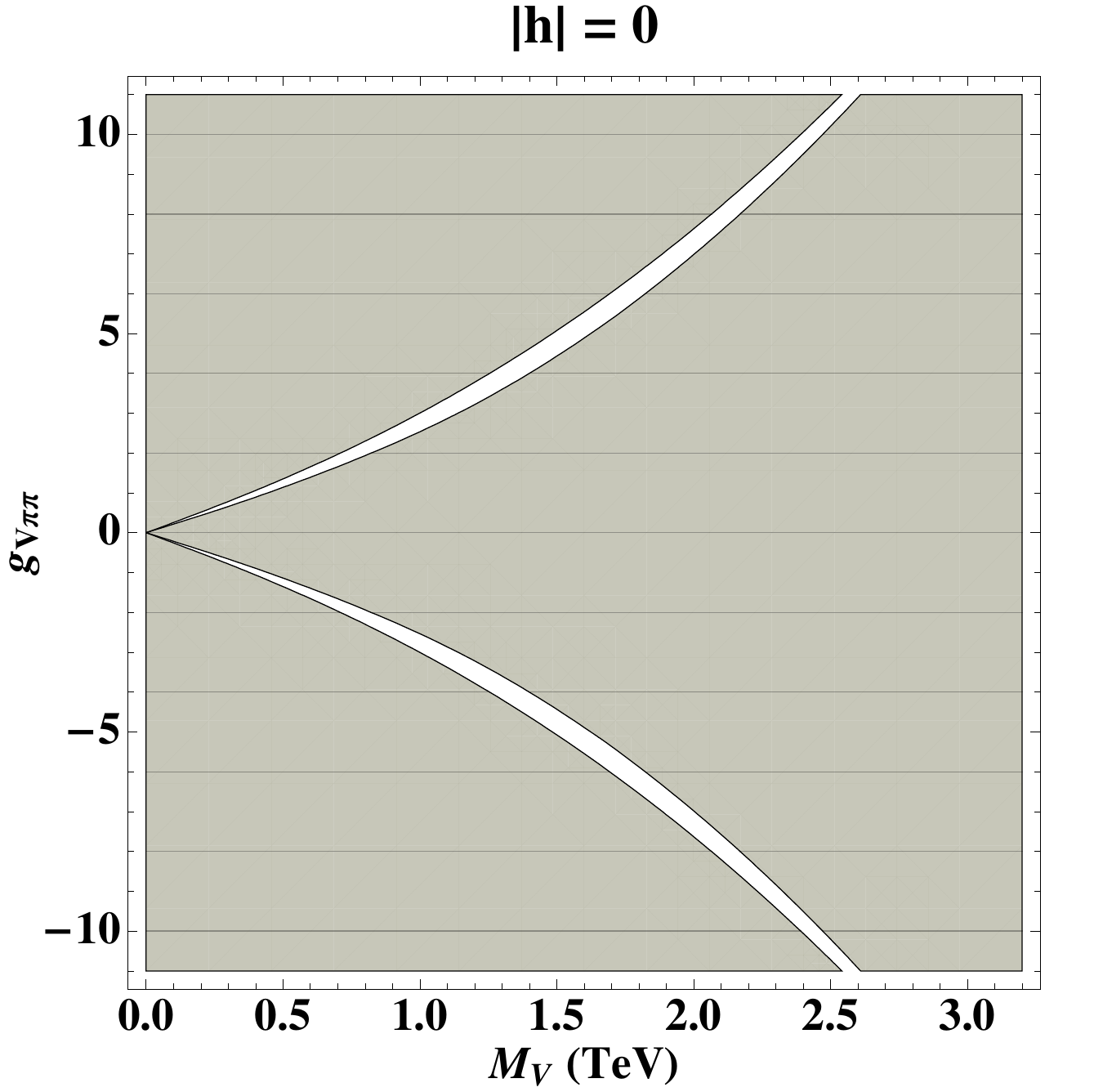}%
\includegraphics[width=0.32\textwidth,height=0.3\textwidth]{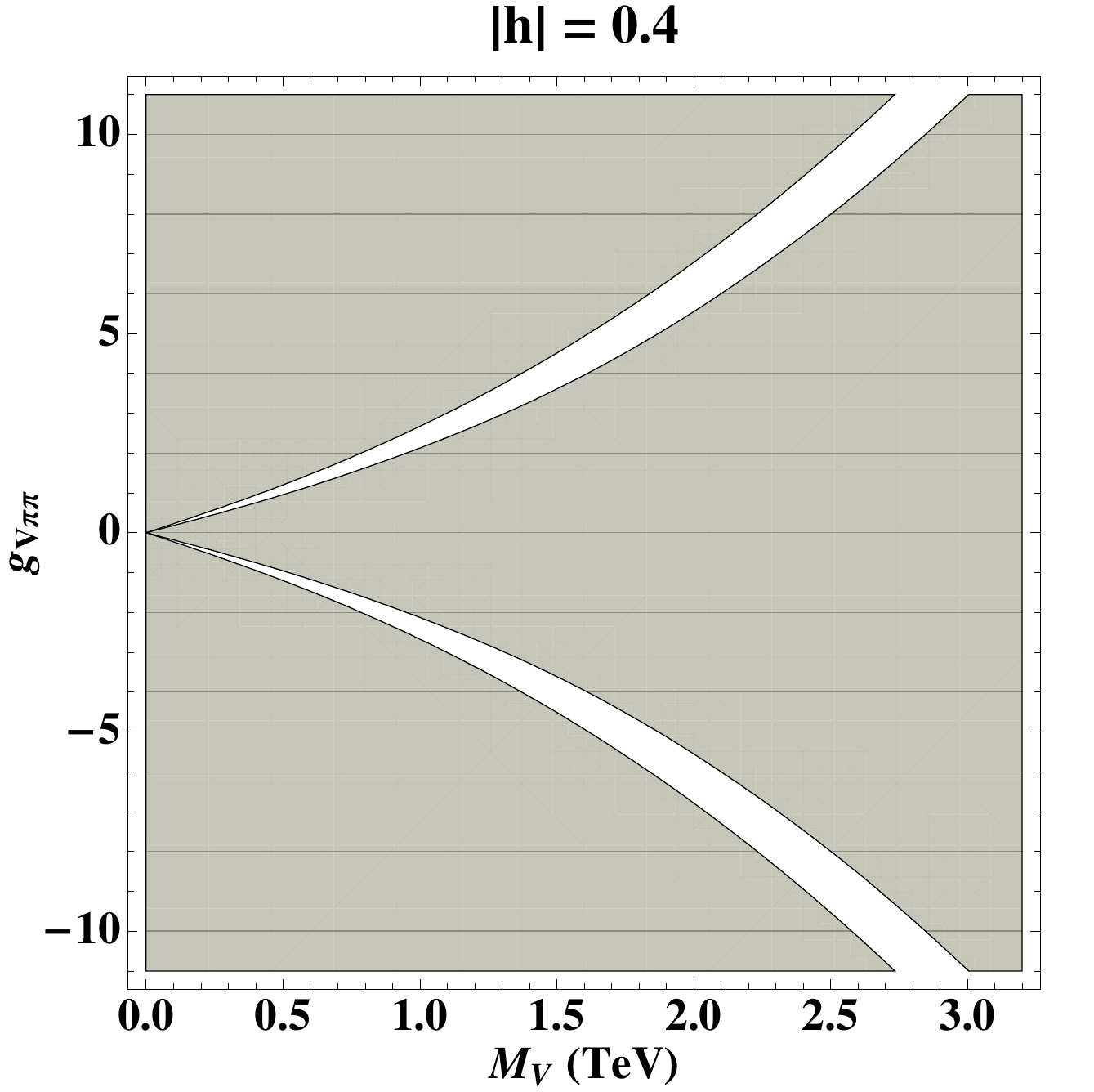}
\includegraphics[width=0.32\textwidth,height=0.3\textwidth]{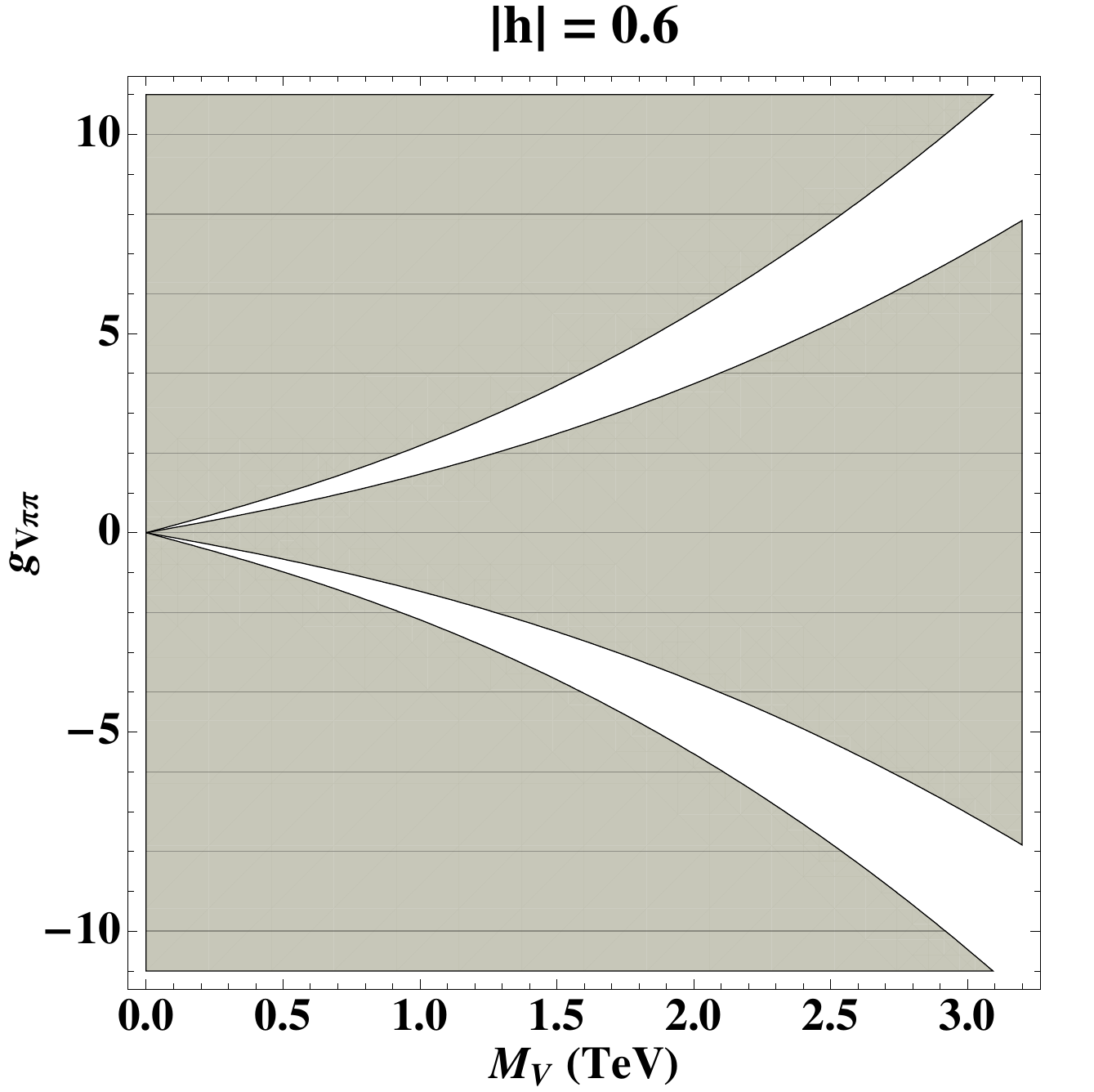}
\includegraphics[width=0.32\textwidth,height=0.3\textwidth]{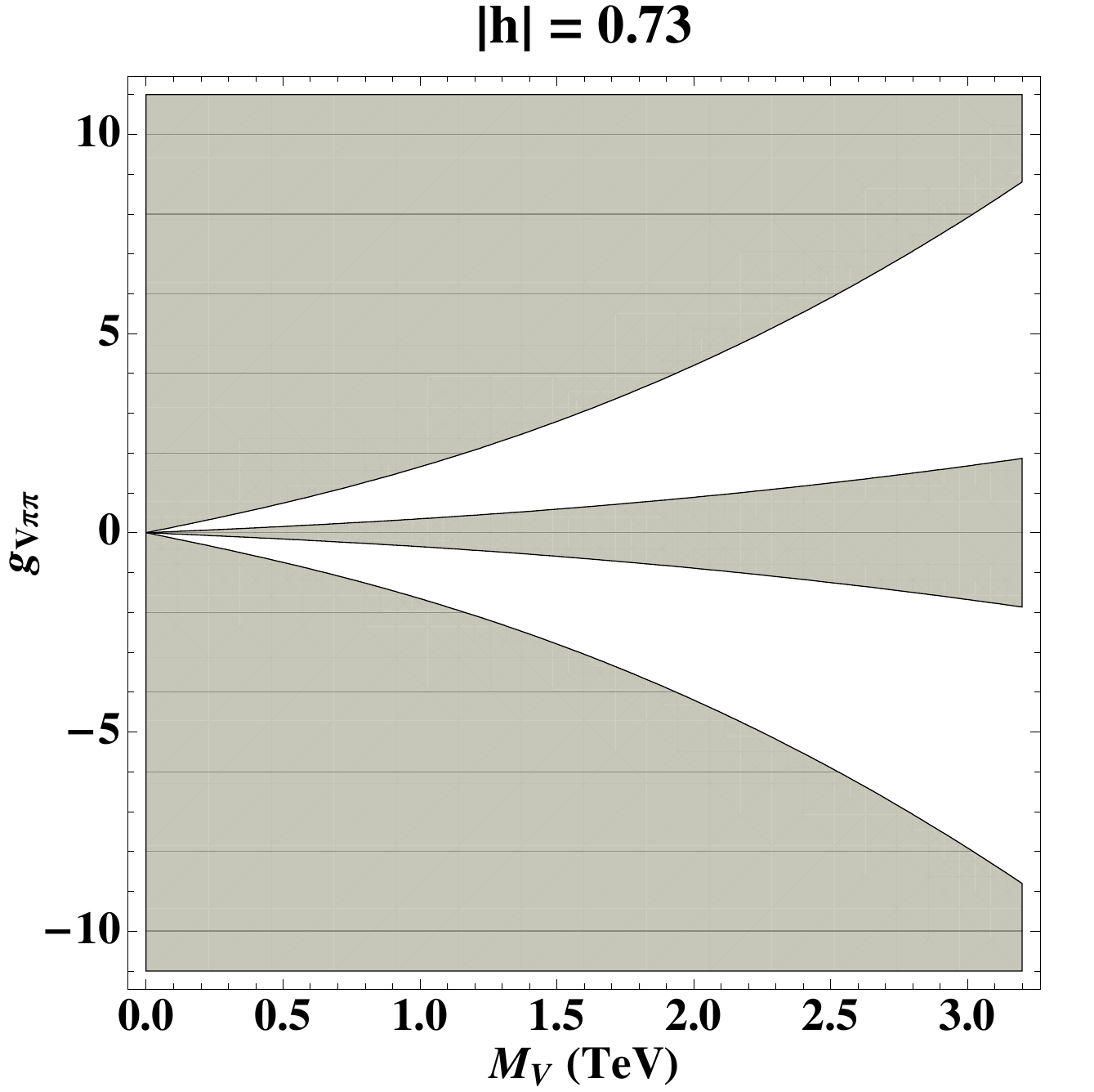}%
\includegraphics[width=0.32\textwidth,height=0.3\textwidth]{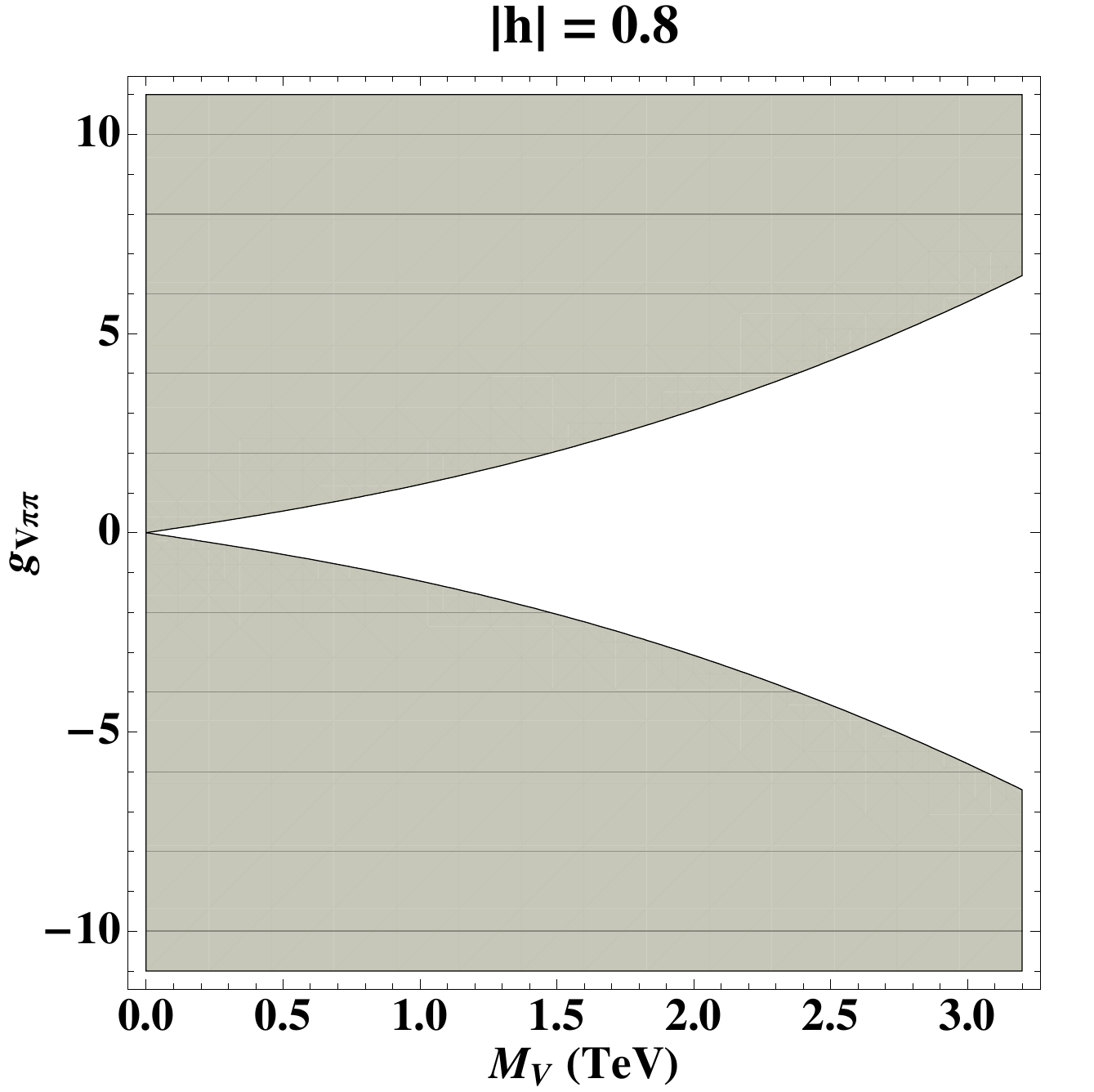}
\includegraphics[width=0.32\textwidth,height=0.3\textwidth]{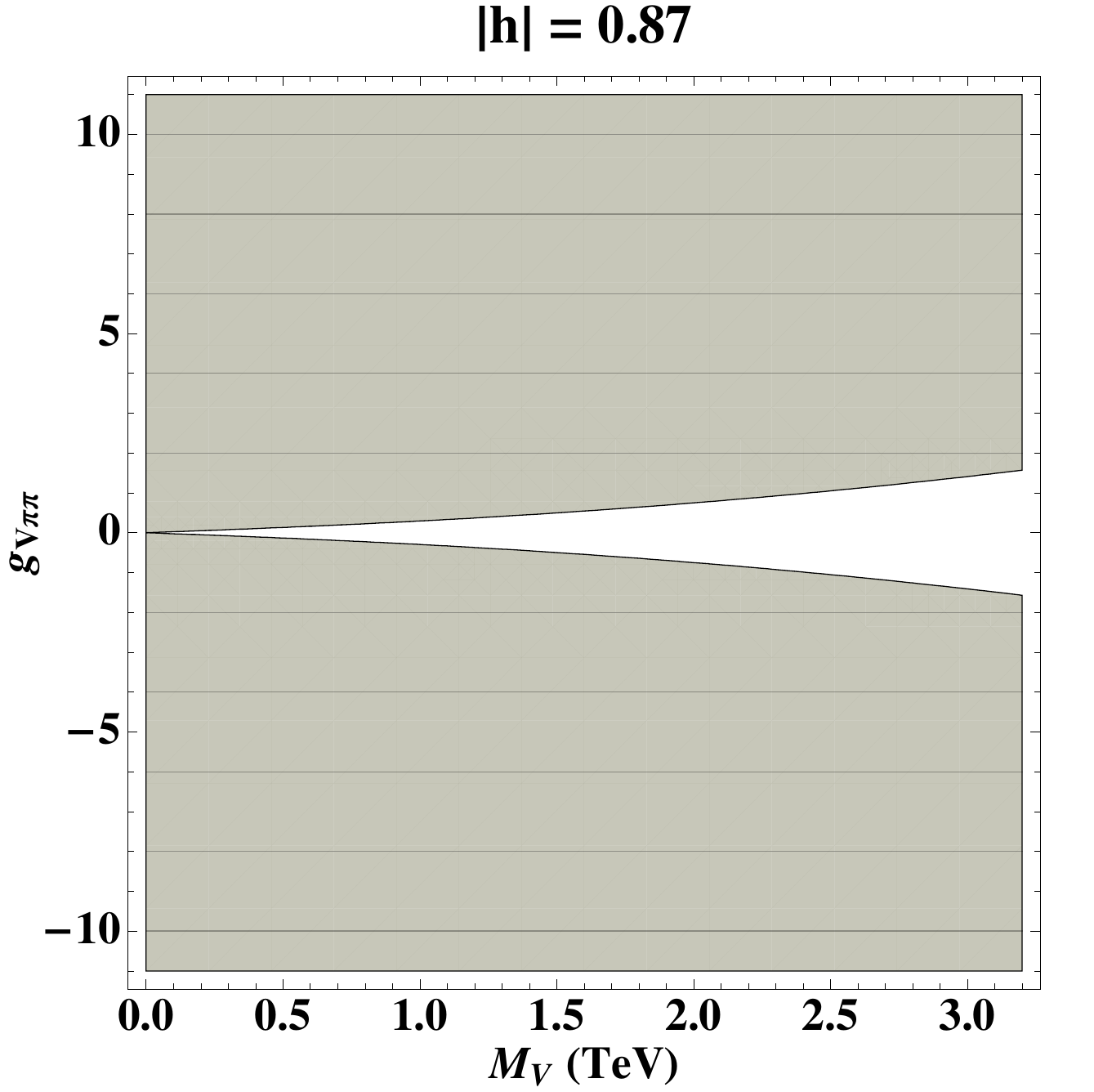}
\caption{Unitarity of $\pi\pi$ scattering up to $\sqrt{s}=3$ TeV in Technicolor with a light Higgs and a spin-one vector resonance. In the white region the $I=0$ $J=0$ partial wave amplitude is within the unitarity bounds, $-1/2\leq a_0^0\leq 1/2$. Different values of the Higgs coupling to pions, $h$, are considered. The $h=0$ case is equivalent to the decoupling limit $M_H\to\infty$, even though for $h=0$ the Higgs only decouples from the pions, while for $M_H\to\infty$ it decouples from the whole theory.}
\label{fig:unit}
\end{figure}

In order to study the effect of the Higgs exchange on the scattering amplitude, consider the high energy behavior of $A(s,t,u)$,
\begin{equation}
A(s,t,u)\sim\left(\frac{1}{F_\pi^2}-\frac{3g_{V\pi\pi}^2}{M_V^2}-\frac{h^2}{M_H^2}\right)s \ .
\label{eq:leading}
\end{equation}
This shows that the Higgs exchange provides an additional negative contribution at large energies, which, together with the vector meson, contributes to delay unitarity violation to higher energies. In Fig.~\ref{fig:a00} $a_0^0$ is plotted as a function of $\sqrt{s}$ for $M_V=1$ TeV, $M_H=200$ GeV, and different values of $g_{V\pi\pi}$ and $h$. The different groups of curves from top to bottom correspond to $g_{V\pi\pi}=$ 2, 2.5, 3, 3.5, and 4. 
For comparison, the QCD value that follows from $\Gamma(\rho\to \pi\pi)\simeq 150$~MeV would be $g_{V\pi\pi}\simeq 5.6$~\footnote{Notice that Fig.~\ref{fig:a00} does not reproduce a scaled up version of QCD $\pi\pi$ scattering. For the latter to occur, the vector resonance should be as large as (246 GeV/93 MeV)$\times$770 MeV $\simeq$ 2 TeV. However in a theory with walking dynamics the resonances are expected to be lighter than in a running setup.}. Within each group, the top curve corresponds to the Higgsless case, $h=0$, while the remaining ones correspond, from top to bottom, to $h=$ 0.1, 0.15, and 0.2. Notice that for small values of $g_{V\pi\pi}$ the presence of a light Higgs delays unitarity violation to higher energies: If the partial wave amplitude has a maximum near 0.5 the delay is dramatic. Notice also that unlike the analysis of Ref.~\cite{Sannino:1995ik} the amplitude zeroes here are not fixed. This is because in Ref.~\cite{Sannino:1995ik} both $g_{V\pi\pi}$ and $F_\pi$ were allowed to scale with the number of colors, while here $F_\pi$ is kept fix at 246 GeV.
 
For a given value of $M_V$, the presence of a light Higgs enlarges the interval of values of $g_{V\pi\pi}$ for which the theory is unitary, provided that $|h|$ is not too large. This is shown in Fig.~\ref{fig:unit}, where the white regions correspond to values of the parameters for which the $I=0$ $J=0$ partial wave amplitude is unitary up to $\sqrt{s}=$ 3 TeV. As $|h|$ grows, the allowed region is enhanced, but as $|h|$ becomes greater than $\simeq 0.9$ the Higgs causes the amplitude to loose unitarity already below $\sqrt{s}=$ 3~TeV regardless of $g_{V\pi\pi}$ and $M_V$. Since the high energy amplitude is dominated by the term linear in $s$, from Eq.~(\ref{eq:leading}) it follows that the corresponding bound is essentially on $h^2/M_H^2$. 

Notice that taking $g_{V\pi\pi}=0$ does not automatically lead to a SM-like behavior of the scattering amplitude. This is most easily seen in the first four plots of Fig.~\ref{fig:unit}, where the $g_{V\pi\pi}=0$ axis lies in a non-unitary region. Indeed,  as shown in Appendix~\ref{appA}, for $g_{V\pi\pi}=0$ the physical pions can still be mixed with the longitudinal component of the axial meson, that is, the parity-even spin-one isotriplet $A^a_\mu$. In order to achieve a true decoupling limit the spin-one resonances should be made infinitely heavy, in which case $g_{V\pi\pi}/M_V\to 0$ and $h\to M_H/F_\pi$, leading to a SM-like unitarization of the $\pi\pi$ scattering amplitude. It is of course true that if $h$ attains the numerical value of $M_H/F_\pi$, then the linear term in $s$ is canceled for $g_{V\pi\pi}=0$, even though the spin-one resonances are not decoupled. For $M_H=200$ GeV this gives $h\to 0.8$. It is therefore expected that the two separate regions of Fig.~\ref{fig:unit} merge at around $|h|\simeq 0.8$.

In this work we focus on theories in which the axial may be lighter than the vector. Due to parity conservation the axial resonance cannot directly participate in the tree-level exchanges of the $\pi\pi$ scattering. As mentioned in the last paragraph the $A^a_\mu$ field appears indirectly in the $\pi\pi$ scattering, since the pion eaten by the $W$ boson contains a certain amount of the longitudinal component of $A^a_\mu$, as Eqs.~(\ref{eq:pion}) and (\ref{eq:FP}) show explicitly. As a consequence the $g_{V\pi\pi}$ and $h$ coupling are affected by the presence of $A^a_\mu$, see Eqs.~(\ref{eq:gVpp}) and (\ref{eq:h_expr}). However the dependence on $M_A$ comes together with new parameters, which make both $g_{V\pi\pi}$ and $h$ completely free to take on any value. The relevant way in which a LAR affects the $\pi\pi$ scattering shows up when the WSR's, together with a small $S$ parameter, are imposed. As we shall see in the next section, this constrains the allowed region in the $(M_V,g_{V\pi\pi})$ plane in a different way than a theory with a QCD-like dynamics and a heavy axial does.

\subsection{Spin-Zero + Spin-One + Spin-Two}

\begin{figure}
\includegraphics[width=0.42\textwidth,height=0.32\textwidth]{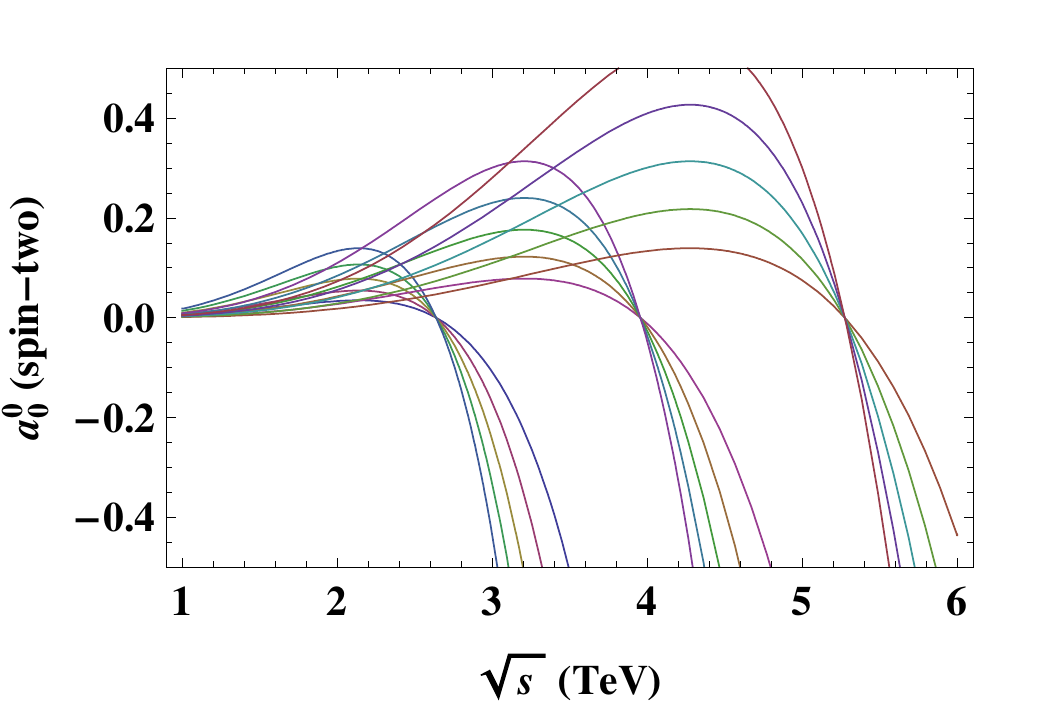}%
\includegraphics[width=0.42\textwidth,height=0.32\textwidth]{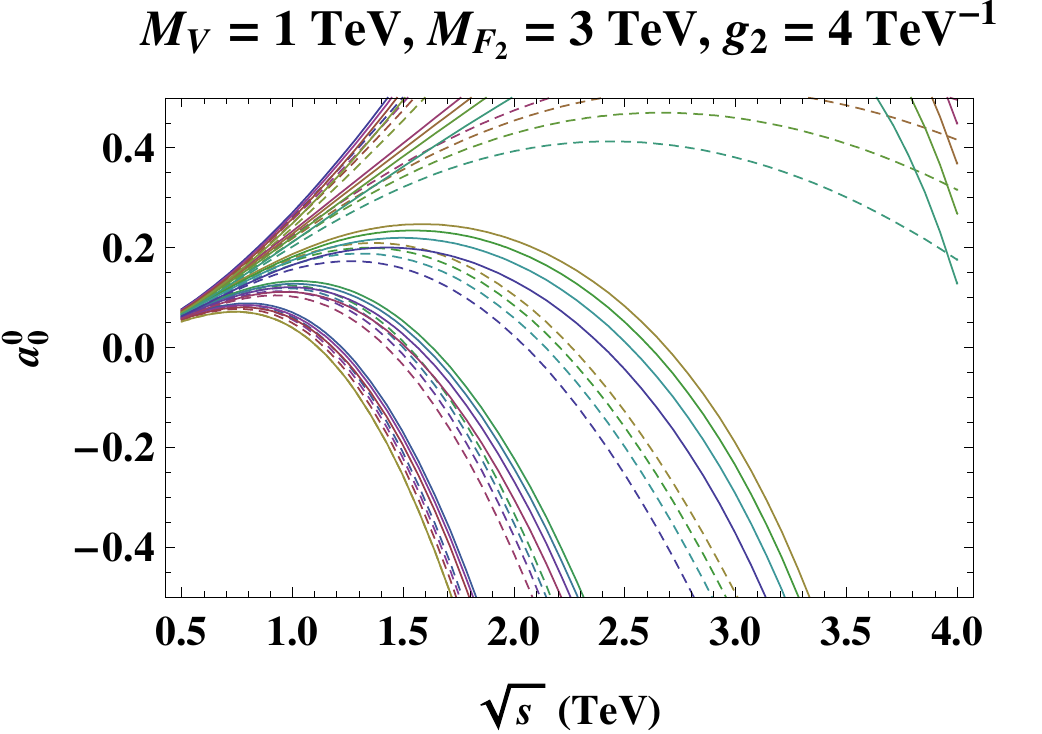}
\caption{Left: Contribution from the spin-two exchanges to the $I=0$ $J=0$ partial wave amplitude of the $\pi\pi$ scattering. The different groups of curves correspond, from left to right, to $M_{F_2}=2,3,4$ TeV. Within each group, the different curves correspond, from smaller to wider, to $g_2=2,2.5,3,3.5,4$~TeV$^{-1}$. Right: $I=0$ $J=0$ partial wave amplitude with all channels included (spin-zero, -one, and -two). The dashed curves reproduce Fig.~\ref{fig:a00}, with just the spin-zero and the spin-one channels included. The solid curves contain also the spin-two exchanges, for $M_{F_2}=3$ TeV, and $g_2=4$~TeV$^{-1}$. If unitarity is violated at negative values of $a_0^0$, the spin-two exchanges may lead to a delay of unitarity violation.}
\label{fig:a00spin2}
\end{figure}

In addition to spin-zero and spin-one mesons, the low energy spectrum can contain spin-two mesons as well~\cite{Harada:1995dc}. The contribution of a spin-two meson $F_2$ to the invariant amplitude is
\begin{eqnarray}
A_2(s,t,u)=\frac{g_2^2}{2(M_{F_2}^2-s)}\left[-\frac{s^2}{3}+\frac{t^2+u^2}{2}\right]-\frac{g_2^2 s^3}{12 M_{F_2}^4} \ ,
\end{eqnarray}
where $M_{F_2}$ and $g_2$ are mass and coupling with the pions, respectively. A reference value for $g_2$ can be obtained from QCD: $m_{f_2}\simeq 1275$~MeV and $\Gamma(f_2\to \pi\pi) \simeq 160$~MeV give $|g_2| \simeq 13$~GeV$^{-1}$ so that $|g_2|F_\pi \simeq 1.2$. Scaling up to the eletroweak scale results in $|g_2| \simeq 4$~TeV$^{-1}$.
The 
contribution of $F_2$ to the $I=0$ $J=0$ partial wave amplitude is given in Fig.~\ref{fig:a00spin2} (left) for different values of $M_{F_2}$ and $g_2$. Notice that the amplitude is initially positive, and then becomes negative at large values of $\sqrt{s}$. If $M_{F_2}$ is large enough, the positive contribution can balance the negative contribution from the spin-zero and spin-one channels, shown in Fig.~\ref{fig:a00}. This can lead to a further delay of unitarity violation, as shown in Fig.~\ref{fig:a00spin2} (right). Here the curves of Fig.~\ref{fig:a00} are redrawn dashed, while the full contribution from spin-zero, spin-one, and spin-two is shown by the solid lines, for $M_{F_2}=3$~TeV and $g_2=4$~TeV$^{-1}$. If unitarity is violated at negative values of $a_0^0$, then the spin-two contibution delays the violation to higher energies.

\section{Unitarity with Walking Dynamics}

The analysis of the previous section was for arbitrary theories with a spin-zero, -one, and -two resonances. However we are mainly interested in analyzing unitarity of $\pi\pi$ scattering in presence of a light Higgs and a LAR, {\em i.e. an axial lighter than the vector}. If VMD holds, the former can only be lighter than the latter in a Walking Technicolor theory (WT), where the second WSR is modified~\cite{Appelquist:1998xf,Foadi:2007ue}. Moreover the chances of the axial being lighter than the vector grow as the conformal window is approached, and the $S$ parameter decreases. Finally, as already mentioned in the Introduction, a LCH can naturally emerge in strongly coupled theories with higher dimensional representations.

Thus in order to consider the LAR scenario we impose the WSR's modified for a WT theory,
\begin{eqnarray}
& & S=4\pi\left[\frac{F_V^2}{M_V^2}-\frac{F_A^2}{M_A^2}\right] \ , \label{eq:WSR0} \\
& & F_V^2-F_A^2=F_\pi^2 \label{eq:WSR1} \ , \\
& & F_V^2 M_V^2 - F_A^2 M_A^2 = a\frac{8\pi^2}{d(R)}F_\pi^4 \label{eq:WSR2} \ ,
\end{eqnarray}
where $F_V$ ($F_A$) and $M_V$ ($M_A$) are decay constant and mass of the vector (axial) resonance, $d(R)$ is the dimension of the fermion representation of the underlying gauge theory, and $a$ is an unknown number, expected to be positive and ${\cal O}(1)$ in WT, and zero in a QCD-like theory. To be more specific, we consider two different gauge theories: Minimal Walking Technicolor (MWT), with two flavors in the adjoint representation of SU(2), and Next-to-Minimal Walking Technicolor (NMWT), with two flavors in the two-index symmetric representation of SU(3). In MWT $d(R)=3$, and the naive contribution to the $S$ parameter is $1/2\pi\simeq 0.15$. As explained in Ref.~\cite{Foadi:2007ue} it is reasonable to take this as a realistic estimate of the full $S$ parameter for this theory. In NMWT $d(R)=6$, and the naive $S$ is $1/\pi\simeq 0.3$. A more recent theory with near conformal dynamics is Ultra Minimal Technicolor: this has the smallest naive contribution to the $S$ parameter, $S=1/3\pi$~\cite{Ryttov:2008xe}. For comparison we also show the constraints for a running theory, {\em i.e.} $a=0$.

The WSR's of Eqs.~(\ref{eq:WSR0})$\div$(\ref{eq:WSR2}) can be generalized to include more vector and axial resonances. It should be noticed however that for the sum rules to hold, these resonances should not be broad. A convenient way to impose this constraint is to exclude regions of the parameter space in which the ratio width/mass is less than a half for both the vector and the axial,
\begin{eqnarray}
\Gamma_V/M_V < 1/2 \ , \quad \Gamma_A/M_A < 1/2 \ .
\label{eq:narrow}
\end{eqnarray}
Formulas for the decay widths are given in Appendix~\ref{appB}.


\begin{figure}
\includegraphics[width=0.32\textwidth,height=0.32\textwidth]{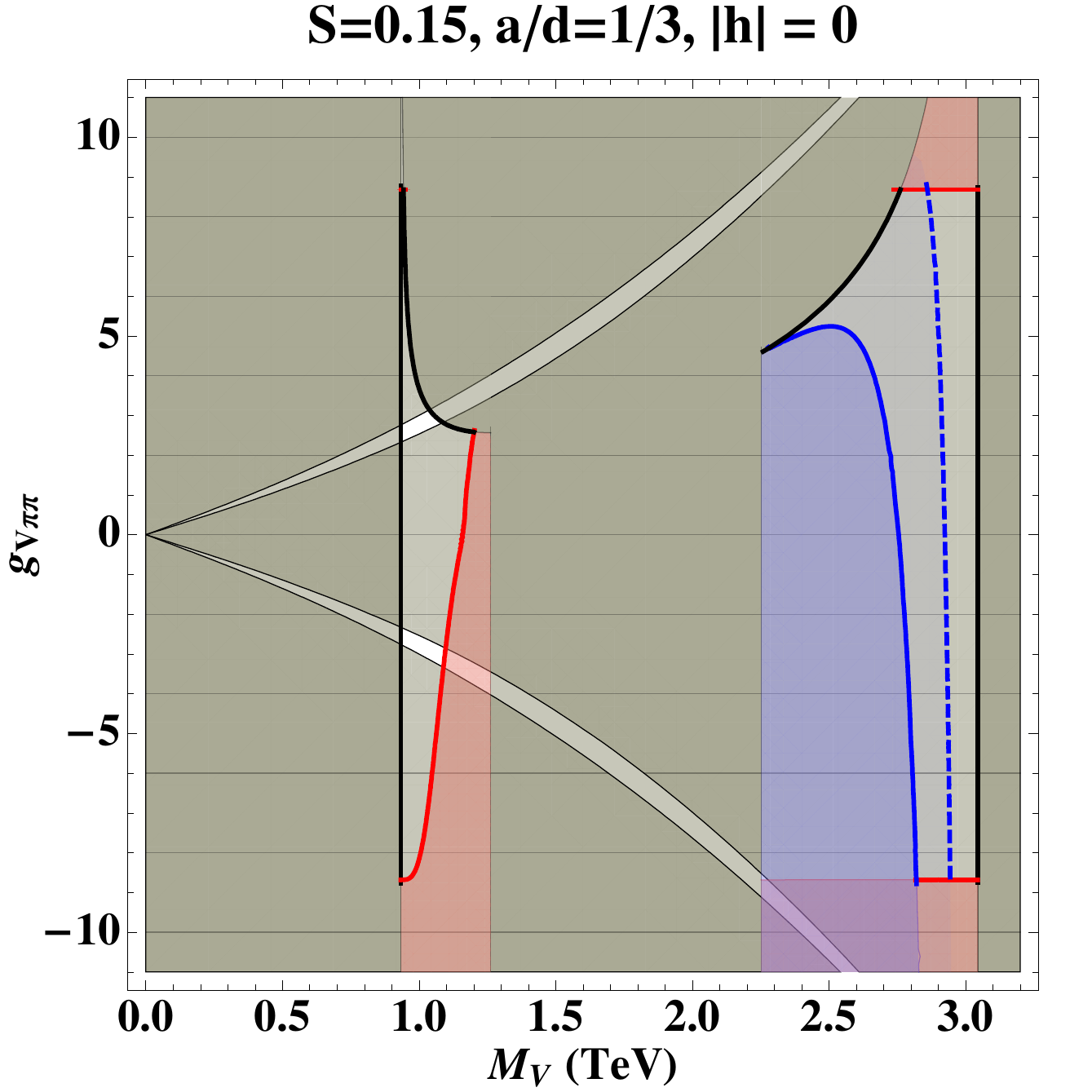}%
\includegraphics[width=0.32\textwidth,height=0.32\textwidth]{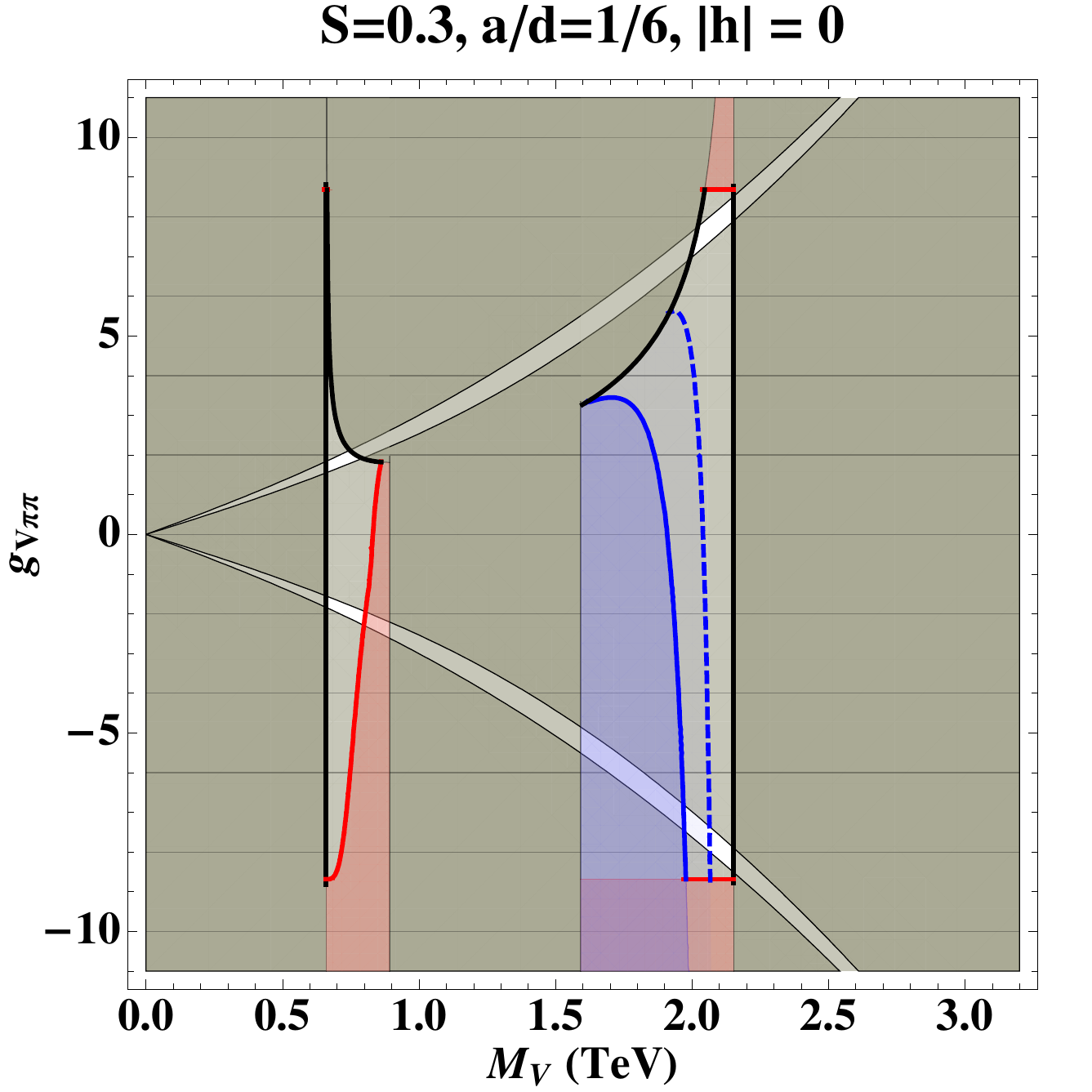}%
\includegraphics[width=0.32\textwidth,height=0.32\textwidth]{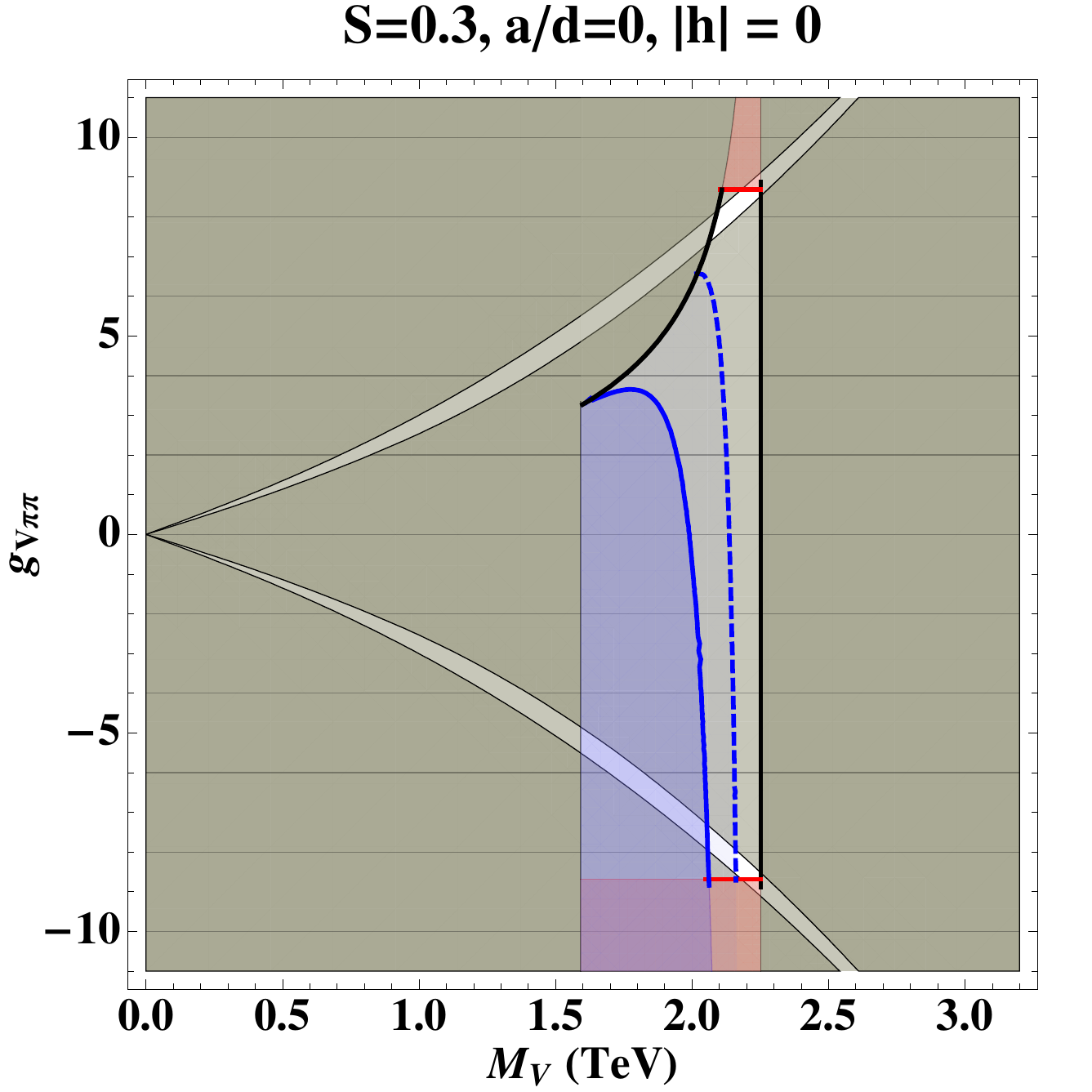}
\vspace{0.1cm}
\includegraphics[width=0.32\textwidth,height=0.32\textwidth]{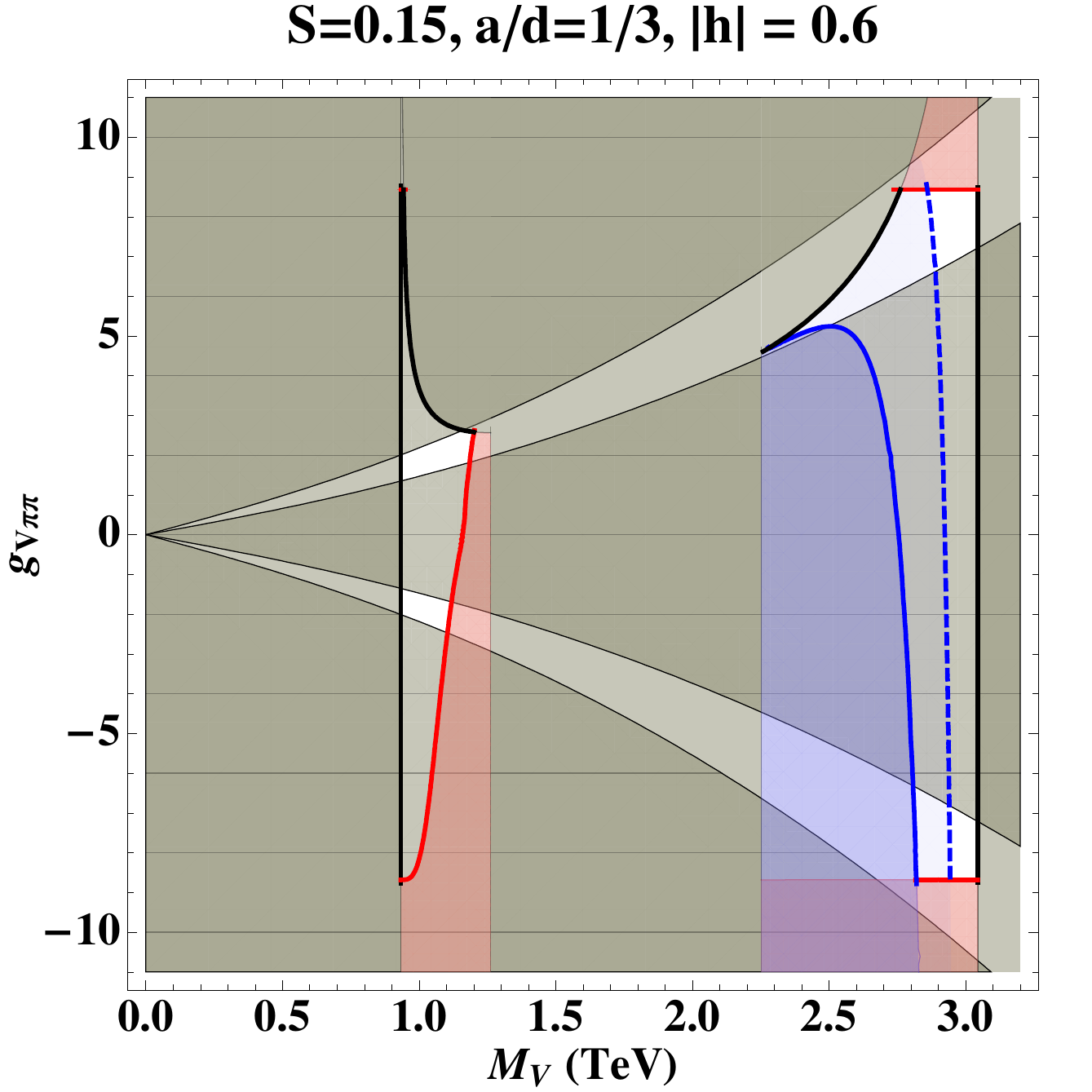}%
\includegraphics[width=0.32\textwidth,height=0.32\textwidth]{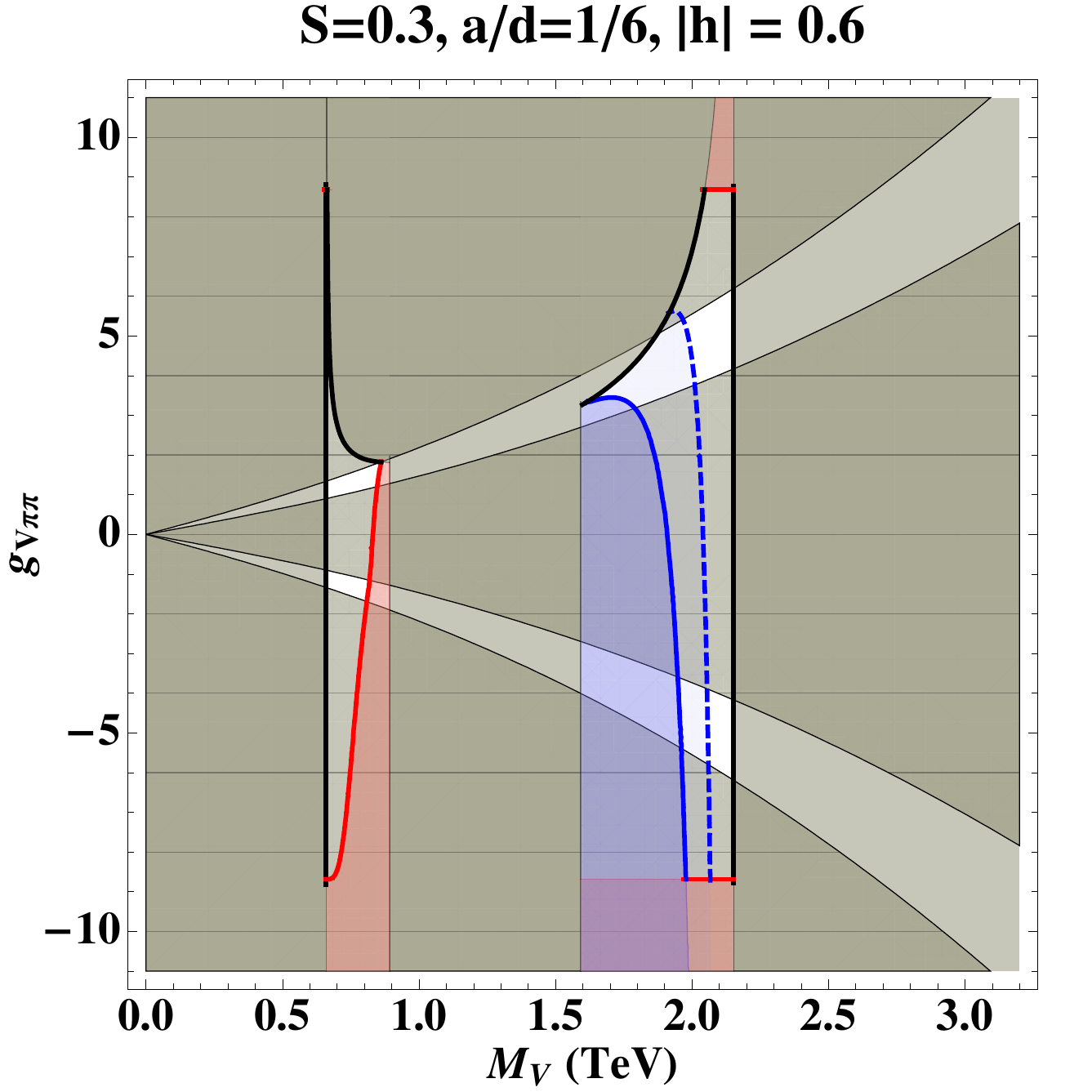}%
\includegraphics[width=0.32\textwidth,height=0.32\textwidth]{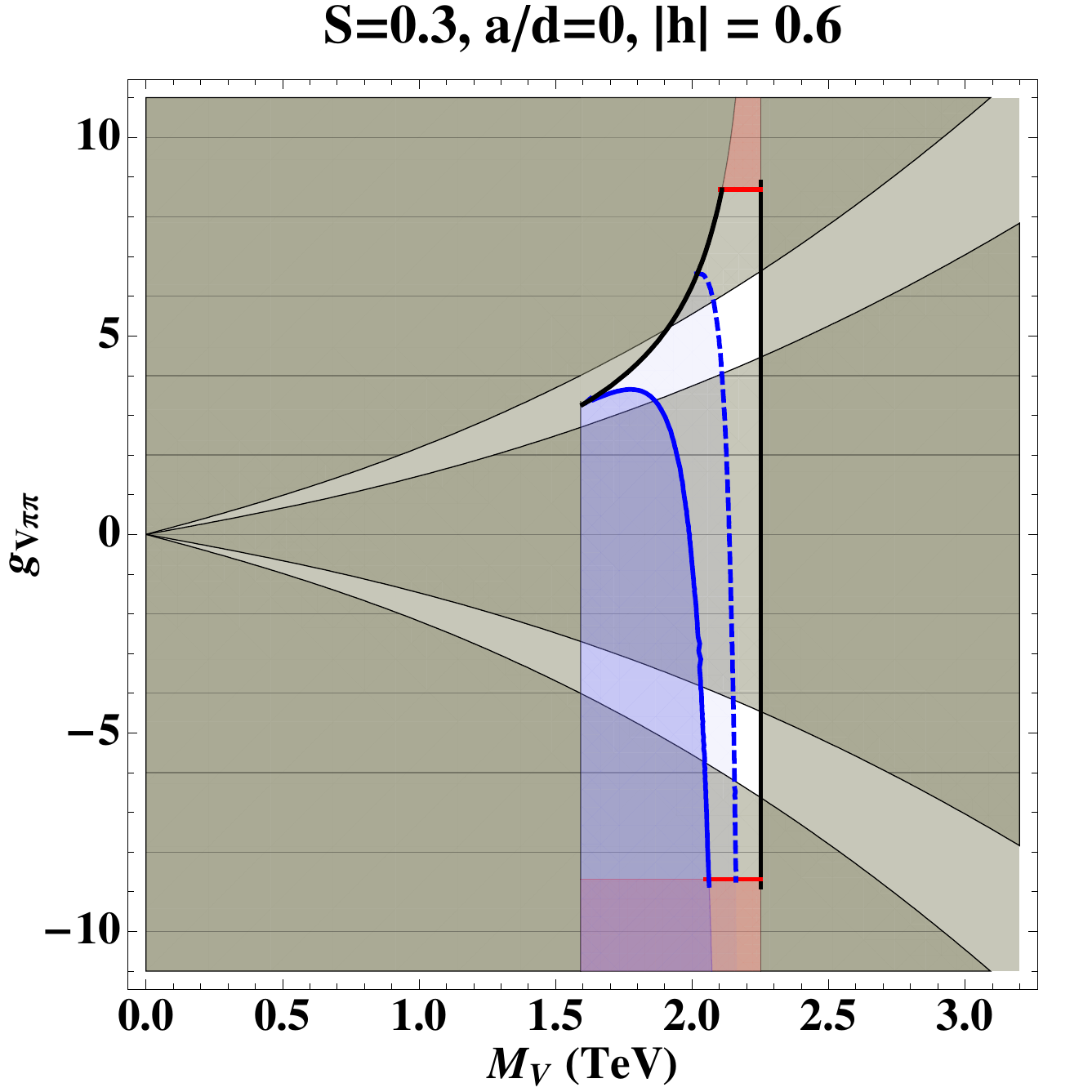}
\vspace{0.1cm}
\includegraphics[width=0.32\textwidth,height=0.32\textwidth]{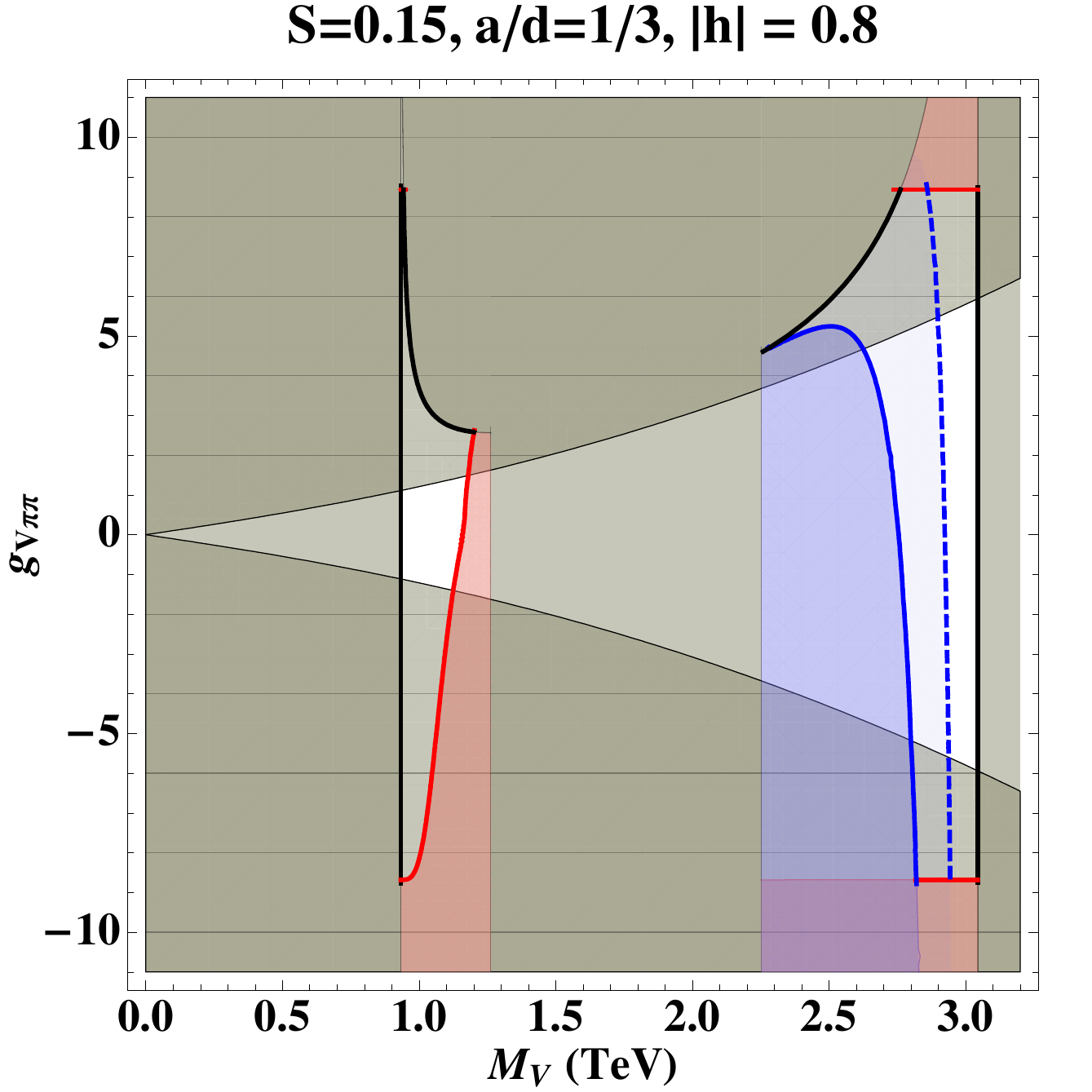}%
\includegraphics[width=0.32\textwidth,height=0.32\textwidth]{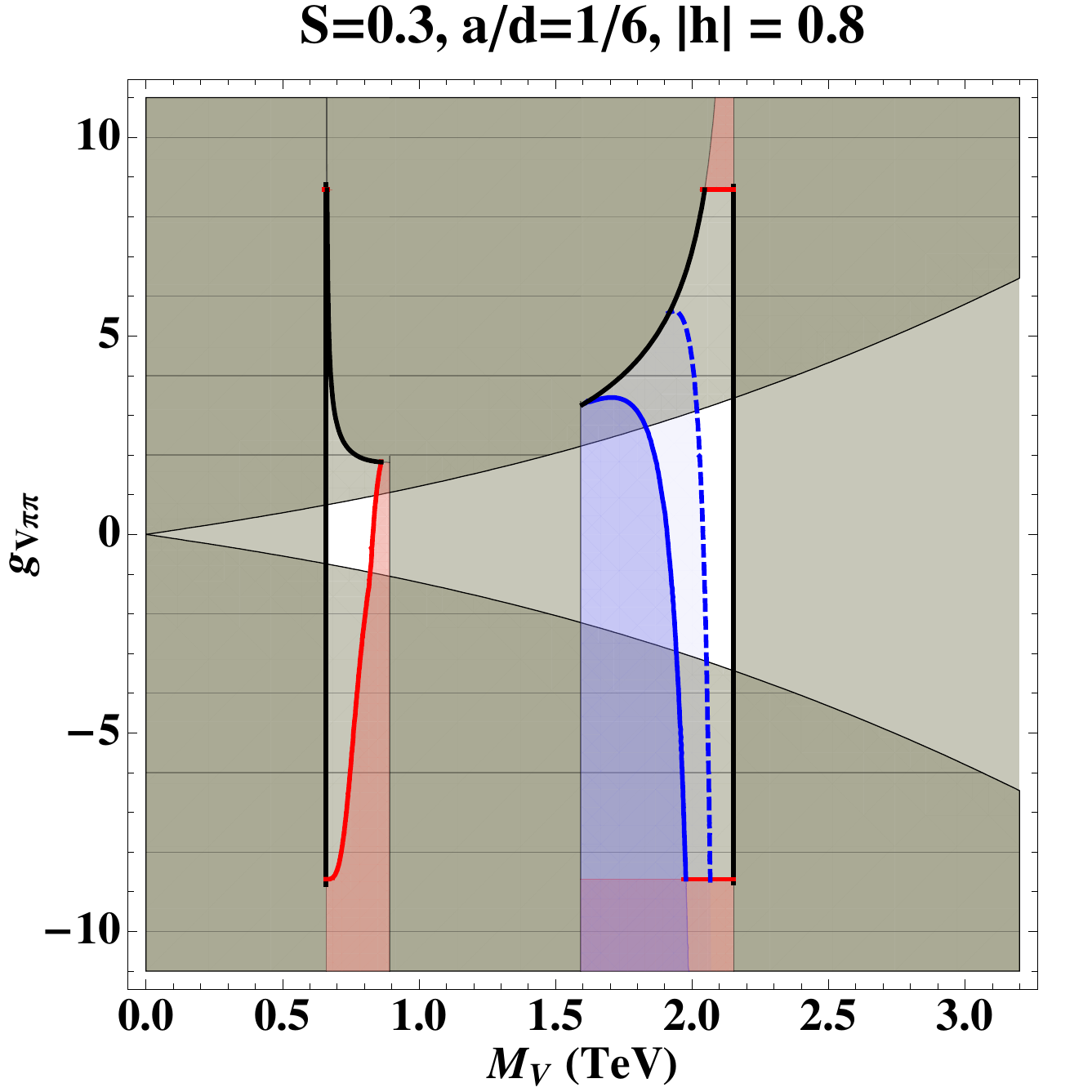}%
\includegraphics[width=0.32\textwidth,height=0.32\textwidth]{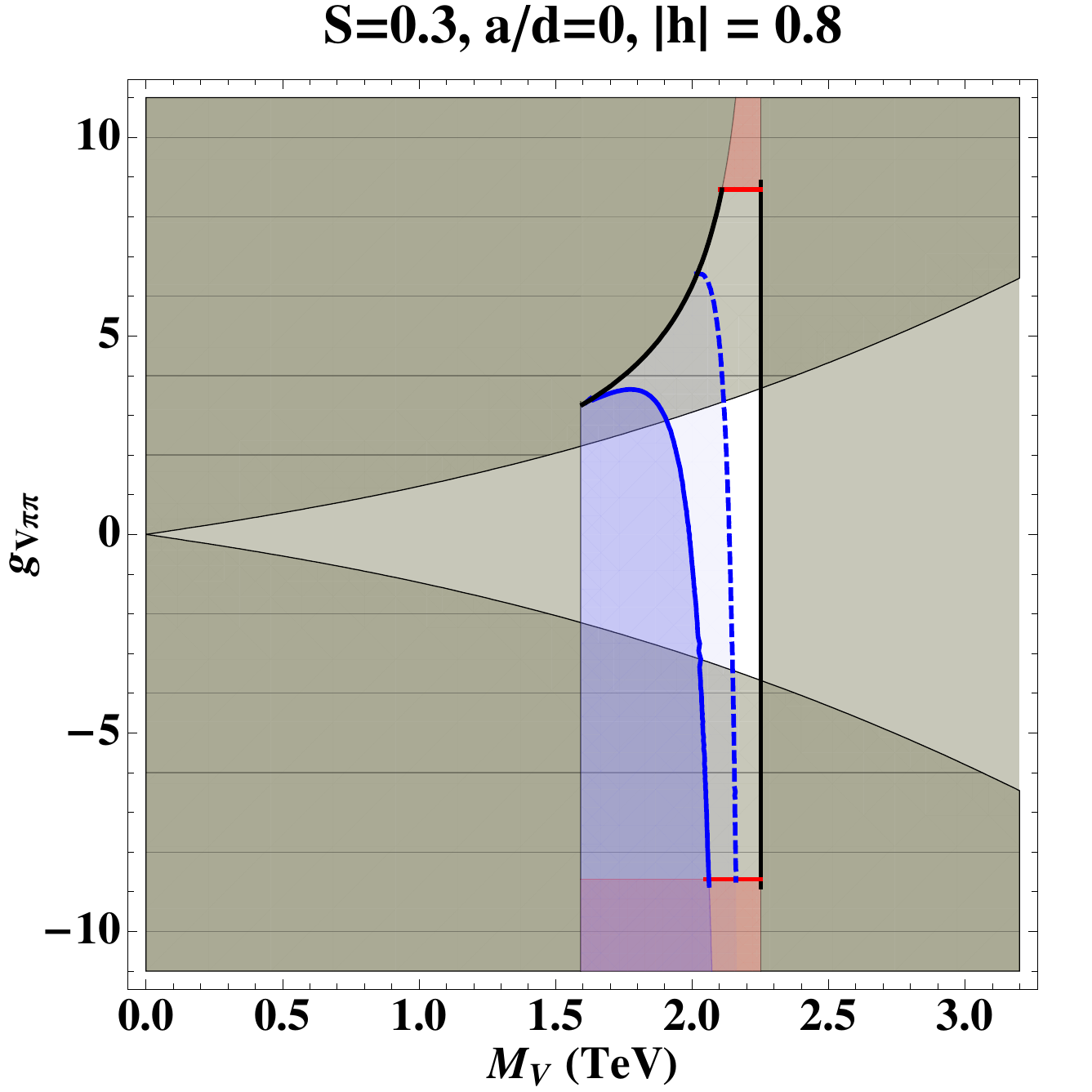}
\vspace{-0.4cm}
\caption{Constraints for $S=0.15$,  $a/d(R)=1/3$ (left), for $S=0.3$, $a/d(R)=1/6$ (center), and $S=0.3$, $a/d(R)=0$ (right). Curves arise from: (i) Unitarity up to $\sqrt{s}=3$~TeV (excluded regions are the striped and shaded ones). (ii) Consistency of the theory (excluded regions are shaded uniformly with gray, located outside the vertical bands and in the upper parts of the bands). (iii) Spin-one vector decay width (excluded regions are the ones shaded uniformly with red in the upper and lower parts of the vertical bands). (iv) Axial decay width (excluded regions are the ones shaded uniformly with blue). We used $M_H=200$~GeV. Thick lines enclose the regions allowed by the constraints (ii)-(iv) while the white regions are allowed by all the contraints.}
\label{fig:combined}
\end{figure}

Integrating these constraints with the unitarity constraints of Fig.~\ref{fig:unit} gives the allowed regions shown in white in Fig.~\ref{fig:combined} (left) for $S=0.15$, $a=1$, $d(R)=3$ (corresponding approximately to MWT), Fig.~\ref{fig:combined} (center) for $S=0.3$, $a=1$, $d(R)=6$ (corresponding approximately to NMWT), and Fig.~\ref{fig:combined} (right) for $S=0.3$, $a=0$ (corresponding approximately to a QCD-like theory)~\footnote{For simplicity in this analysis we ignore the contribution from the spin-two resonance.}. In particular, the vertical bands are the only regions in which the WSR's of Eqs.~(\ref{eq:WSR0})-(\ref{eq:WSR2}) can be satisfied. The left band is determined by Eq.~(\ref{eq:MVlim}) in Appendix~\ref{appC}, and disappears for $a=0$. In this band the axial is lighter than the vector. The right band is determined by Eq.~(\ref{eq:MVlim2}) and is still present for $a=0$. In this band the axial is heavier than the vector. Above the uppermost curve within each band the theory exhibits tachyonic states (see Eq.~(\ref{eq:gVpplim})), and the corresponding regions are therefore excluded. 

The top and bottom horizontal (red) lines in the right band, together with the lower curve on the left band, come from the requirement $\Gamma_V/M_V<1/2$. The lower (blue) curve on the right band comes from the requirement $\Gamma_A/M_A<1/2$ for $|g_{AH\pi}-h_{AH\pi}|=0$, where the couplings $g_{AH\pi}$ and $h_{AH\pi}$ are defined in the Appendix, and parametrize the strength of the $A\to H,\pi$ decay. The thick closed curves enclose the regions allowed by all the constraints except unitarity. The white region is allowed by all the constraints in each of the plots. The requirement $\Gamma_A/M_A<1/2$ depends on $|g_{AH\pi}-h_{AH\pi}|$ and the Higgs mass. For $M_H=200$ GeV, values of $|g_{AH\pi}-h_{AH\pi}|$ above $\sim 17$ give no allowed regions in the heavy regime. The thick dashed curve in the left band shows how this constraint is altered for $|g_{AH\pi}-h_{AH\pi}|=17$.

From Fig.~\ref{fig:combined} we see that imposing the modified WSR's together with a small $S$ parameter, and demanding unitarity of the $\pi\pi$ scattering implies:
\begin{itemize}
\item Unitarity without a Higgs at $\sqrt{s}=  3$~TeV is only possible in a restricted region of the parameter space.
\item In presence of a LAR, the $\pi\pi$ scattering can be unitary at $\sqrt{s}=  3$~TeV even without a Higgs and for small values of $S$. This is in agreement with the results of Ref.~\cite{Foadi:2008ci}. Of course the reason for this is that in this case also the vector resonance is forced to be light, and can therefore unitarize the amplitude. 
\item For larger values of $S$ the vector meson masses become smaller, and both regimes move to smaller values of $M_V$. This makes unitarity without a Higgs possible even with a heavy axial.
\item In presence of a LAR unitarity demands $g_{V\pi\pi}$ to be small, even if a light Higgs is in the spectrum.
\item With a heavy axial, unitarity demands $g_{V\pi\pi}$ to be large either in the Higgsless scenario, or if the coupling of the Higgs to the pions is not sufficiently large.
\item With a light Higgs and a suitable value of the coupling to the pions, most of the region that is allowed by the other constraints can be unitarized up to $\sqrt{s}=  3$~TeV, both in the light and the heavy meson regime. As the $|g_{AH\pi}-h_{AH\pi}|$ coupling is increased, the heavy meson regime becomes less and less viable for narrow axial resonances, but the theory is still unitary in a large portion of the
parameter space.
\item In a QCD-like theory a LAR is not allowed by the constraints imposed by the traditional WSR's. Therefore in a QCD-like theory $g_{V\pi\pi}$ is expected to be large.
\end{itemize}

\section{Conclusions}

In this paper we have analyzed the $WW$ scattering in Technicolor models with near confromal dynamics, in which both a 200 GeV LCH and a LAR are in the low energy spectrum. As expected, the LCH significantly enlarges the parameter space in which the tree-level $\pi\pi$ scattering is unitary at a certain scattering energy (which has been chosen to be $\sqrt{s}=3$ TeV), provided that its coupling $h$ to the pion is neither too small nor too large. 

A LAR affects the analysis on the $\pi\pi$ scattering by imposing constraints on certain regions of the parameter space which are not constrained by unitarity. The constraints are imposed through the modified WSR's, for certain specific gauge theories (namely MWT and NMWT). In order for the WSR's to hold the spin-one resonances should be narrow: this imposes further constraints on the parametre space. Our analysis shows that in presence of a LAR $g_{V\pi\pi}$ is required to be small, regardless of the LCH. Furthermore unitarity in a Higgsless theory is possible with a LAR, even for small values of the $S$ parameter, since in this case also the vector resonance is forced to be light~\cite{Foadi:2008ci}. 

WT is also compatible with a heavy axial resonance. In this scenario, which is the only possible for a QCD-like Technicolor, the Higgsless $\pi\pi$ scattering demands a large $g_{V\pi\pi}$ and a large $S$ parameter, while the Higgsful $\pi\pi$ scattering is unitary at 3 TeV in a very large portion of the available parameter space, provided that the coupling $h$ is within certain bounds.

Finally we remind the reader that the present analysis can be extended to include broad resonance effects, in which case the amplitude for the $\pi\pi$ scattering cannot be fully perturbative, and some unitarization schemes must be employed.

\acknowledgments
We are thankful to D.D. Dietrich, M.T. Frandsen, and J. Schechter for useful discussions.

\appendix

\section{Lagrangian and Vertices \label{appA}}

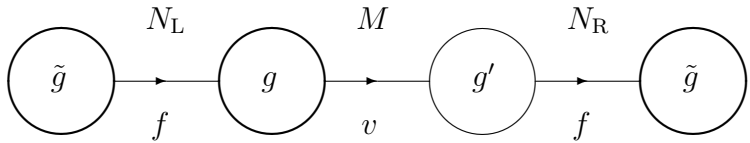
\begin{figure}
\vspace{-5cm}
\setlength{\unitlength}{7cm}
\begin{center}
\begin{picture}(1,1)
\put(0.1,0){\vector(1,0){0.1}}
\put(0.2,0){\line(1,0){0.1}}
\put(0.5,0){\vector(1,0){0.1}}
\put(0.6,0){\line(1,0){0.1}}
\put(0.9,0){\vector(1,0){0.1}}
\put(1.0,0){\line(1,0){0.1}}
\thicklines
\put(0,0){\circle{0.2}}
\put(0.4,0){\circle{0.2}}
\thinlines
\put(0.8,0){\circle{0.2}}
\thicklines
\put(1.2,0){\circle{0.2}}
\put(-0.02,-0.01){$\tilde{g}$}
\put(0.38,-0.01){$g$}
\put(0.78,-0.01){$g^\prime$}
\put(1.18,-0.01){$\tilde{g}$}
\put(0.17,-0.1){$f$}
\put(0.57,-0.1){$v$}
\put(0.97,-0.1){$f$}
\put(0.16,0.1){$N_{\rm L}$}
\put(0.56,0.1){$M$}
\put(0.96,0.1){$N_{\rm R}$}
\end{picture}
\end{center}
\vspace{1cm}
\caption{Moose diagram for a chiral resonance model with a spontaneously broken SU(2)$_{\rm L}\times$SU(2)$_{\rm R}$ chiral symmetry. Each circle represents an SU(2) global symmetry. In the thick circles the full SU(2) symmetry is gauged, in the thin circle only the U(1) subgroup is gauged. The two circles at the ends of the chain correspond to the vector mesons, while the internal circles correspond to the ordinary SM gauge group, which is a subgroup of the chiral symmetry group. $N_{\rm L}$ and $N_{\rm R}$ are nonlinear sigma fields, with VEV $f$. Since a light Higgs is included in the spectrum, $M$ is taken to be a linear sigma field, with VEV $v$}
\label{fig:moose}
\end{figure}
Technicolor theories with an SU(2)$_{\rm L}\times$SU(2)$_{\rm R}$ chiral symmetry and vector resonances can be described by promoting the latter to gauge fields ${A^a_{\rm L}}_\mu$ and ${A^a_{\rm R}}_\mu$ of a mirror gauge group SU(2)$_{\rm L}^\prime\times$SU(2)$_{\rm R}^\prime$. The full symmetry group is then SU(2)$_{\rm L}^\prime\times$SU(2)$_{\rm L}\times$SU(2)$_{\rm R}$SU(2)$_{\rm R}^\prime$, where the electroweak bosons $\widetilde{W}^a_\mu$ and $\widetilde{B}_\mu$ are the gauge fields of SU(2)$_{\rm L}$ and the U(1) subgroup of SU(2)$_{\rm R}$. This model can be described by the four-site moose diagram of Fig.~\ref{fig:moose}. The vector fields acquire their ``hard'' mass through the SU(2)$_{\rm L}^\prime\times$SU(2)$_{\rm L}\to$SU(2)$_{\rm L, diag}$ and SU(2)$_{\rm R}\times$SU(2)$_{\rm R}^\prime\to$SU(2)$_{\rm R, diag}$ symmetry breaking mechanisms. ``Before'' chiral symmetry breaking this model contains massless $\widetilde{W}^a_\mu$ and $\widetilde{B}_\mu$ fields, together with massive vector resonances, all transforming under an unbroken SU(2)$_{\rm L, diag}\times$SU(2)$_{\rm R, diag}$ symmetry. The very fact that this chiral symmetry group is different from the original SU(2)$_{\rm L}\times$SU(2)$_{\rm R}$ one, in absence of vector fields, already shows that the latter do affect the chiral dynamics.

The model contains nonlinear sigma fields $N_{\rm L}$ and $N_{\rm R}$, 
\begin{eqnarray}
N_{\rm L}=\exp\left(2\ i\ \widetilde{\pi}^a_{\rm L}\ T^a\ /\ f\right)\ , \quad 
N_{\rm R}=\exp\left(2\ i\ \widetilde{\pi}^a_{\rm R}\ T^a\ /\ f\right) \ ,
\end{eqnarray}
and a linear sigma field $M$,
\begin{eqnarray}
M=\frac{1}{\sqrt{2}}\left(v+H+2\ i\ \widetilde{\pi}^a\ T^a\right) \ ,
\end{eqnarray}
where $T^a=2\tau^a$, and $\tau^a$ are the Pauli matrices. Here $\pi^a_{\rm L}$ and $\pi^a_{\rm R}$ are the pions produced in the SU(2)$_{\rm L}^\prime\times$SU(2)$_{\rm L}\to$SU(2)$_{\rm L, diag}$ and SU(2)$_{\rm R}\times$SU(2)$_{\rm R}^\prime\to$SU(2)$_{\rm R, diag}$ symmetry breaking mechanisms, respectively, with vacuum expectation value (VEV) $f$, while $\pi^a$ are the pions produced in SU(2)$_{\rm L}\times$SU(2)$_{\rm R}\to$SU(2)$_{\rm V}$, with VEV $v$. $H$ is of course the composite Higgs.

Assuming, in the limit of decoupled spin-one mesons, a SM-like Higgs sector, the ${\cal O}(p^2)$ Lagrangian invariant under the parity transformations
\begin{eqnarray}
{A_{\rm L}}_\mu\to {A_{\rm R}}_\mu\ , \quad N_{\rm L}\to N_{\rm R}^\dagger\ , \quad M\to M^\dagger
\end{eqnarray}
can be written as
\begin{eqnarray}
{\cal L} & = & -\frac{1}{2}{\rm Tr}\left[\widetilde{W}_{\mu\nu}\widetilde{W}^{\mu\nu}\right]
-\frac{1}{4}\widetilde{B}_{\mu\nu}\widetilde{B}^{\mu\nu} \nonumber \\
& - &\frac{\kappa(\xi)}{2}{\rm Tr}\left[{F_{\rm L}}_{\mu\nu}F_{\rm L}^{\mu\nu}+{F_{\rm R}}_{\mu\nu}F_{\rm R}^{\mu\nu}\right]
-\frac{2\gamma(\xi)}{f^2}
{\rm Tr}\left[N_{\rm L}^\dagger {F_{\rm L}}_{\mu\nu} N_{\rm L} M N_{\rm R} {F_{\rm R}}_{\mu\nu} N_{\rm R}^\dagger M^\dagger\right] \nonumber \\
& + & \frac{f^2 k(\xi)}{4}{\rm Tr}\left[D_\mu N_{\rm L}^\dagger D^\mu N_{\rm L} + D_\mu N_{\rm R}^\dagger D^\mu N_{\rm R} \right]
+\frac{1}{2}{\rm Tr}\left[D_\mu M^\dagger D^\mu M\right] \nonumber \\
&+& r_2(\xi){\rm Tr}\left[D_\mu N_{\rm L}^\dagger N_{\rm L} M D^\mu N_{\rm R} N_{\rm R}^\dagger M^\dagger\right]
+ \frac{r_3(\xi)}{4}{\rm Tr}\left[D_\mu N_{\rm L}^\dagger N_{\rm L}\left(M D^\mu M^\dagger-D_\mu M M^\dagger\right)\right. \nonumber \\
& + & \left. D^\mu N_{\rm R} N_{\rm R}^\dagger \left(M^\dagger D^\mu M - D_\mu M^\dagger M\right)\right] -{\cal V}(M) \ ,
\label{eq:Lagr}
\end{eqnarray}
where
\begin{equation}
\xi\equiv \frac{1}{v^2}{\rm Tr}[M M^\dagger] \ ,
\end{equation}
and the potential can be expanded to quartic order to be
\begin{eqnarray}
{\cal V}(M)=-\frac{v^2\lambda}{2} {\rm Tr}[M M^\dagger] + \frac{\lambda}{4} {\rm Tr}[M M^\dagger]^2 \ .
\end{eqnarray}

The covariant derivatives are
\begin{eqnarray}
D_\mu M&=&\partial_\mu M -i\ g\ \widetilde{W}_\mu^a\ T^a M + i\ g^\prime \ M\ B_\mu\ T^3 \nonumber \\
D_\mu N_{\rm L}&=&\partial_\mu N_{\rm L}-i\ \tilde{g}\ A_{{\rm L}\mu}^a\ T^a\ N_{\rm L}
+i\ g\ N_{\rm L}\ \widetilde{W}_\mu^a\ T^a \nonumber \\
D_\mu N_{\rm R}&=&\partial_\mu N_{\rm R}-i\ g^\prime\ B_\mu\ T^3\ N_{\rm R}
+i\ \tilde{g}\ N_{\rm R}\ A_{{\rm R}\mu}^a\ T^a \ .
\end{eqnarray}
The analytic functions $\kappa(\xi)$, $\gamma(\xi)$, $k(\xi)$, $r_2(\xi)$, $r_3(\xi)$ are arbitrary~\footnote{The terminology here refers to Ref.~\cite{Foadi:2007ue}, where the chiral symmetry is SU(4), and an additional $r_1$ term is present, and $r_1=r_2=r_3=0$ in the Custodial Technicolor limit.  In SU(N)$_{\rm L}\times$SU(N)$_{\rm R}$ such term can be absorbed in $k$.}, and should be expanded around the VEV $\xi=1$.

Here we are mainly interested in the strongly interacting sector, which can be obtained by switching off the electroweak couplings, $g,g^\prime\to 0$. When this is done, the canonically normalized vector and axial resonances are found to be
\begin{eqnarray}
V^a_\mu & = & \sqrt{1 + \frac{v^2\gamma(1)}{f^2}}\ \frac{{A^a_{\rm L}}_\mu+{A^a_{\rm R}}_\mu}{\sqrt{2}} \ , \nonumber \\
A^a_\mu & = & \sqrt{1 - \frac{v^2\gamma(1)}{f^2}}\ \frac{{A^a_{\rm L}}_\mu-{A^a_{\rm R}}_\mu}{\sqrt{2}} \ ,
\end{eqnarray}
with masses
\begin{eqnarray}
M_{V}^2 & = & \frac{g_V^2}{4}\ \left[f^2-r_2(1)v^2\right] \ , \label{eq:MV} \\
M_{A}^2 & = & \frac{g_A^2}{4}\ \left[f^2+r_2(1)v^2\right] \ , \label{eq:MA}
\end{eqnarray}
where the couplings to the vector and the axial, $g_V$ and $g_A$, respectively, are
\begin{eqnarray}
g_V\equiv \frac{\tilde{g}}{\sqrt{1 + \displaystyle{\frac{v^2\gamma(1)}{f^2}}}} \ , \quad 
g_A\equiv \frac{\tilde{g}}{\sqrt{1 - \displaystyle{\frac{v^2\gamma(1)}{f^2}}}} \ .
\end{eqnarray}
The decay constants are
\begin{eqnarray}
F_V & = & \frac{\sqrt{2}M_V}{g_V} \ , \label{eq:FV} \\
F_A & = & \frac{\sqrt{2}M_A}{g_A}
\left[1-\frac{r_3(1)\ g_A^2\ v^2}{4 M_A^2}\right] \label{eq:FA} \ .
\end{eqnarray}

The longitudinal components of $V^a$ and $A^a$ are the canonically normalized eaten pions
\begin{eqnarray}
\pi^a_V & = & \frac{2\ M_V}{g_V\ f} \ \frac{\widetilde{\pi}^a_{\rm L}-\widetilde{\pi}^a_{\rm R}}{\sqrt{2}} \ , \nonumber \\
\pi^a_A & = & \frac{2\ M_A}{g_A\ f} \ \frac{\widetilde{\pi}^a_{\rm L}
+\widetilde{\pi}^a_{\rm R}}{\sqrt{2}}+\sqrt{1-\frac{F_\pi^2}{v^2}} \ \widetilde{\pi}^a \ ,
\end{eqnarray}
while the remaining orthogonal combination is the canonically normalized physical pion, eaten by the SM gauge bosons when the electroweak couplings are switched on:
\begin{eqnarray}
\pi^a & = & \frac{F_\pi}{v} \ \widetilde{\pi}^a \ .
\label{eq:pion}
\end{eqnarray}
Here $F_\pi$ is $\pi^a$ decay constant,
\begin{eqnarray}
F_\pi = v \sqrt{1-\frac{r_3^2(1)\ g_A^2\ v^2}{8 M_A^2}} \ .
\label{eq:FP}
\end{eqnarray}
In a Technicolor theory $F_\pi\simeq 246$ GeV. Notice that $v\geq F_\pi$, and $F_\pi\to v$ as the longitudinal component of the axial decouples from the pion. This occurs if either $M_A\to\infty$ or $r_3\to 0$.

When expanded in terms of the physical fields, Eq.~(\ref{eq:Lagr}) gives the Lagrangian terms
\begin{eqnarray}
{\cal L}_{V\pi\pi} & = & g_{V\pi\pi}\ \varepsilon^{abc}\ V^a_\mu\ \pi^b\ \partial^\mu\pi^c \ , \label{eq:LVpp} \\
{\cal L}_{AV\pi} & = & g_{AV\pi}\ F_\pi\ \varepsilon^{abc}\ V^a_\mu\ {A^b}^\mu\ \pi^c
+ h_{AV\pi}\ F_\pi\ \varepsilon^{abc}\ V^a_{\mu\nu}\ {A^b}^{\mu\nu}\ \pi^c \ , \\
{\cal L}_{AAV} & = & g_{AAV}\ \varepsilon^{abc}\ V_{\mu\nu}^a\ {A^b}^\mu\ {A^c}^\nu
+h_{AAV}\ \varepsilon^{abc}\ A_{\mu\nu}^a\ {A^b}^\mu\ {V^c}^\nu \ , \\
{\cal L}_{AH\pi} & = & g_{AH\pi}\ H\ A_\mu^a\ \partial^\mu\pi^a
+ h_{AH\pi}\ \partial^\mu H\ A_\mu^a\ \pi^a \ , \\
{\cal L}_{H\pi\pi} & = & h_1\ M_H\ H\ \pi^a\ \pi^a + \frac{h_2}{F_\pi}\ H\ \partial^\mu\pi^a\ \partial_\mu\pi^a
+ \frac{h_3}{F_\pi}\ \partial^\mu H \partial_\mu\pi^a\ \pi^a \ , \label{eq:LHpp} \\
{\cal L}_{\pi\pi\pi\pi} & = & g_1\ \pi^a\ \pi^a\ \pi^b\ \pi^b + \frac{g_2}{F_\pi^2}\ \pi^a\ \pi^a\ \partial^\mu\pi^b\ \partial_\mu\pi^b
+ \frac{g_3}{F_\pi^2}\ \pi^a\ \partial^\mu\pi^a\ \pi^b\ \partial_\mu\pi^b \ , \label{eq:Lpppp}
\end{eqnarray}
plus other quartic terms which are not relevant for our analysis. The couplings are found to be
\begin{eqnarray}
g_{V\pi\pi} & = & \frac{F_V M_V}{2\ F_\pi^2}\left(1-\frac{g_A^2 F_A^2}{2 M_A^2}\right)  \ , \label{eq:gVpp} \\
g_{AV\pi} & = & \frac{M_A M_V}{F_\pi^2}\frac{F_A}{F_V}\left(1-\frac{g_A^2}{g_V^2}\frac{M_V^2}{M_A^2}\right) \ , \\
h_{AV\pi} & = & \frac{1}{2}\frac{M_V}{M_A}\frac{F_A}{F_V}\left(\frac{g_A^2}{g_V^2}-1\right) \ , \\
g_{AAV} & = & \frac{1}{2}\frac{g_A^2}{g_V^2}\frac{M_V}{F_V} \ , \\
h_{AAV} & = & \frac{M_V}{F_V} \ , \\
g_{AH\pi} & = & -\frac{g_A\ f^2}{\sqrt{2}\ F_\pi\ v}\left(1-\frac{g_A F_A}{\sqrt{2} M_A}\right)
\left[\left(1+\frac{g_V^2}{g_A^2}\frac{M_A^2}{M_V^2}\right)^{-1}
-k^\prime(1)-\frac{\ r_2^\prime(1)\ v^2}{f^2}\right] \nonumber \\
&-&\frac{r_3^\prime(1)\ g_A\ v}{\sqrt{2}\ F_\pi} \ , \\
h_{AH\pi} & = & -\frac{F_A\ M_A}{v\ F_\pi}\left(1-\frac{g_A F_A}{\sqrt{2} M_A}\right)  \ , \\
h_1 & = & -\frac{v\ M_H}{2\ F_\pi^2} \ , \\
h_2 & = & -\frac{g_A^2\ f^2\ F_\pi}{4\ v\ M_A^2}\left(\frac{v^2}{F_\pi^2}-1\right)
\left[1-k^\prime(1)-\frac{\ r_2^\prime(1)\ v^2}{f^2}\right] \nonumber \\
&-&\frac{r_3^\prime(1)\ g_A\ v\ \sqrt{v^2-F_\pi^2}}{\sqrt{2}\ M_A\ F_\pi} \ , \\
h_3 & = & \frac{1}{v}\left(\frac{v^2}{F_\pi^2}-1\right) \ , \\
g_1 & = & -\frac{v^2\ M_H^2}{8\ F_\pi^4} \ , \\
g_2 & = & -\frac{F_V^2}{8F_\pi^2}\left(1-\frac{g_A\ F_A}{\sqrt{2}\ M_A}\right)^2
\Bigg[1+2\ \frac{g_V^2}{g_A^2}\frac{M_A^2}{M_V^2}-\frac{\sqrt{2}\ F_A\ g_A}{M_A}
\left(1+\frac{g_A\ F_A}{2\sqrt{2}\ M_A}\right) \nonumber \\
&-& 2\ k^\prime(1)\left(1+\frac{g_V^2}{g_A^2}\frac{M_A^2}{M_V^2}\right)
\Bigg]+\frac{v^2}{2\ F_\pi^2}
\Bigg[- r_3^\prime(1)\left(1-\frac{g_A\ F_A}{\sqrt{2}\ M_A}\right)
\nonumber \\
&+&\frac{r_2^\prime(1)}{2}\left(1-\frac{g_A\ F_A}{\sqrt{2}\ M_A}\right)^2\Bigg] \ , \\
g_3 & = & -\frac{F_V^2}{8F_\pi^2}\left(1-\frac{g_A\ F_A}{\sqrt{2}\ M_A}\right)^2
\left[1-4\ \frac{g_V^2}{g_A^2}\frac{M_A^2}{M_V^2}+\frac{\sqrt{2}\ F_A\ g_A}{M_A}
\left(1+\frac{g_A\ F_A}{2\sqrt{2}\ M_A}\right)\right] \ .
\end{eqnarray}

The coupling $h$ defined in Eq.~(\ref{eq:h}) is then
\begin{eqnarray}
h & = & -\frac{M_H}{v}\Bigg[1-\frac{g_A^2\ f^2}{4\ M_A^2}\left(\frac{v^2}{F_\pi^2}-1\right)
\left(1-k^\prime(1)-\frac{r_2^\prime(1)\ v^2}{f^2}\right) \nonumber \\
&-&\frac{r_3^\prime(1)\ g_A\ v^2\ \sqrt{v^2-F_\pi^2}}{\sqrt{2}\ M_A\ F_\pi^2}\Bigg] \ .
\label{eq:h_expr}
\end{eqnarray}

A few words should be said about the decoupling limits. The spin-one mesons can be decoupled from the pions and the Higgs by letting $f\to\infty$, as shown by Eqs.(\ref{eq:MV}), (\ref{eq:MA}). If this occurs the Higgs-pion system becomes identical to the SM one, as the equations above show explicitly. For the axial to be decoupled alone, {\em i.e.} without the vector, one must have $\gamma\to f^2/v^2$, since in this case $g_A\to\infty$ and $g_V$ stays finite. When this occurs the Higgs-pion system differs from the SM one, because of the presence of the vector resonance. Finally notice that setting $g_{V\pi\pi}\to 0$ does not necessarily lead to a SM $h$ coupling. In fact when $g_{V\pi\pi}\to 0$ the spin-one resonances are still there to mix with the Higgs-pion system.

\section{Decay Widths}\label{appB}

The spin-one meson decay channels are $V\to\pi,\pi$, $V\to A,\pi$, $V\to A,A$ for the vector, and $A\to V,\pi$, $A\to H\pi$ for the axial. The partial decay widths are
\begin{eqnarray}
& & \Gamma_{V\to\pi\pi} = \frac{g_{V\pi\pi}^2 M_V}{48\pi}\left(1-\frac{4 M_\pi^2}{M_V^2}\right)^{3/2} \label{eq:GVpp} \ ,  \\
& & \Gamma_{V\to A\pi} =  \frac{\sqrt{\lambda(M_V^2,M_A^2,M_\pi^2)}}{24\pi M_V^3}
\Bigg[g_{AV\pi}^2\left(3+\frac{\lambda(M_V^2,M_A^2,M_\pi^2)}{4 M_V^2 M_A^2}\right) \nonumber \\
& + & 6 g_{AV\pi} h_{AV\pi} \left(M_V^2+M_A^2-M_\pi^2\right)
+2h_{AV\pi}^2\left(\lambda(M_V^2,M_A^2,M_\pi^2)+6 M_V^2 M_A^2\right)\Bigg] \ , \\
& & \Gamma_{V\to A A} = \frac{M_V}{48 M_A^4 \pi }\left(1-\frac{4M_A^2}{M_V^2}\right)^{3/2}
\bigg[g_{VAA}^2 M_V^4 + \nonumber \\
& &  \left(4
   g_{AAV}^2+6 h_{AAV}
   g_{AAV}+h_{AAV}^2\right) M_V^2
   M_A^2+3 g_{AAV}^2 M_A^4\bigg] \ , \\
& & \Gamma_{A\to V\pi} = \frac{\sqrt{\lambda(M_V^2,M_A^2,M_\pi^2)}}{24\pi M_A^3}
\Bigg[g_{AV\pi}^2\left(3+\frac{\lambda(M_V^2,M_A^2,M_\pi^2)}{4M_V^2 M_A^2}\right) \nonumber \\
& + & 6 g_{AV\pi} h_{AV\pi} \left(M_V^2+M_A^2-M_\pi^2\right)
+2h_{AV\pi}^2\left(\lambda(M_V^2,M_A^2,M_\pi^2)+6 M_V^2 M_A^2\right)\Bigg] \ , \\
& & \Gamma_{A\to H\pi} = (g_{AH\pi}-h_{AH\pi})^2\frac{\lambda(M_A^2,M_H^2,M_\pi^2)^{3/2}}{192\pi M_A^5} \ ,
\end{eqnarray}
where
\begin{eqnarray}
\lambda(x,y,z) \equiv x^2+y^2+z^2-2 x y - 2 y z - 2 z x \ .
\end{eqnarray}

\section{Constraints from the WSR's}\label{appC}

Taking $F_\pi$, $S$, $a$, and the $g_{AH\pi}$ coupling as input, Eqs.~(\ref{eq:WSR0})$\div$(\ref{eq:WSR2}) and Eq.~(\ref{eq:narrow}) impose constraints on the parameter space $(M_V,g_{V\pi\pi})$. Eqs.~(\ref{eq:WSR0})$\div$(\ref{eq:WSR2}) give
\begin{eqnarray}
& & F_A^2 = \frac{1-\displaystyle{\frac{2\pi a S}{d(R)}}}{1-\displaystyle{\frac{M_V^2 S}{8\pi F_\pi^2}}-\displaystyle{\frac{4\pi^2 a F_\pi^2}{d(R) M_V^2}}}\frac{F_\pi^2}{2}-F_\pi^2 > 0 \ , \label{eq:FAS} \\
& & M_A^2 = \frac{1-\displaystyle{\frac{8 a \pi^2 F_\pi^2}{d(R)M_V^2}}}{\displaystyle{\frac{M_V^2 S}{4\pi F_\pi^2}}-1}M_V^2 > 0 \ , \label{eq:MAS}
\end{eqnarray}
which in turn imply the inequalities
\begin{eqnarray} \label{eq:MVlim}
\frac{4\pi}{S}\left(1-\sqrt{1-\frac{2\pi a\, S}{d(R)}}\right) < \frac{M_V^2}{F_\pi^2}  <   \frac{8 \pi^2 a}{d(R)} \ , \\
 \label{eq:MVlim2}
\frac{4\pi}{S} <  \frac{M_V^2}{F_\pi^2}   <  \frac{4\pi}{S}\left(1+\sqrt{1-\frac{2\pi a\, S}{d(R)}}\right) \ .
\end{eqnarray}
Eq.~(\ref{eq:GVpp}) gives
\begin{eqnarray}
g_A^2 = \frac{2M_A^2}{F_A^2}\left[1-\frac{2 F_\pi^2 g_{V\pi\pi}}{F_V M_V}\right] > 0 \ , \label{eq:gA}
\end{eqnarray}
which implies the bound
\begin{eqnarray}
g_{V\pi\pi} < \frac{F_V M_V}{2 F_\pi^2} \ , \label{eq:gVpplim}
\end{eqnarray}
where $F_V=\sqrt{F_A^2+F_\pi^2}$, and $F_A$ is given by Eq.~(\ref{eq:FAS}). This last inequality must be satisfied in order to prevent tachyonic states from showing up in the theory.

\end{document}